\documentclass[english,notitlepage]{revtex4-2}
\usepackage[T1]{fontenc}
\usepackage[utf8]{inputenc}
\usepackage{colortbl}
\usepackage{babel}
\usepackage{amsmath}
\usepackage{amssymb}
\usepackage{graphicx}
\usepackage[bookmarks=false,
 breaklinks=false,pdfborder={0 0 1},backref=false,colorlinks=false]
 {hyperref}

\makeatletter

\providecommand{\tabularnewline}{\\}

\usepackage{colortbl}
\usepackage{babel}

\usepackage{slashed}

\providecommand{\tabularnewline}{\\}

\usepackage{xcolor}
\usepackage{pdfcolmk}
\usepackage{ulem}
\usepackage{lineno} 

\providecolor{lyxadded}{rgb}{0,0,1}
\providecolor{lyxdeleted}{rgb}{1,0,0}

\DeclareRobustCommand{\lyxsout}[1]{\ifx\\#1\else\sout{#1}\fi}

\newcommand{\Mmin}{M_{\mathrm{min}}}

\newcommand{\Gpre}{\mathcal{G}}
\allowdisplaybreaks

\makeatother

\begin{document}
\title{Production of quarkonium pairs via the fragmentation mechanism}
\author{Franco Barattini, Benjamin Guiot, Marat Siddikov }
\affiliation{Departamento de Física, Universidad Técnica Federico Santa María,
~\\
 y Centro Científico-Tecnológico de Valparaíso, Casilla 110-V, Valparaíso,
Chile}
\begin{abstract}
In this manuscript, we analyze the contribution of fragmentation mechanisms
to the inclusive hadroproduction of heavy quarkonium pairs, taking
into account both single- and double-hadron (dihadron) fragmentation
channels. To estimate the contribution of the latter mechanism, we
construct a microscopic perturbative model of the dihadron fragmentation
function, which is valid in the limit of large invariant masses of
the heavy quarkonium pair. Using the Color Glass Condensate framework
to evaluate the $Q\bar{Q}$ production amplitude, we find that in
LHC kinematics, the single fragmentation of heavy quarks into quarkonium
is only a minor correction, whereas dihadron fragmentation can provide
a sizable contribution, on par with the contributions of single- and
double-parton scattering. 
\end{abstract}
\maketitle

\section{Introduction}

Due to their large masses, heavy quarkonia are widely used as important
probes of the gluonic field of the target~\citep{Korner:1991kf,Neubert:1993mb}.
The nonrelativistic QCD (NRQCD) framework allows us to systematically
evaluate various perturbative corrections to processes involving quarkonium~\citep{Bodwin:1994jh,Maltoni:1997pt,Brambilla:2008zg,Feng:2015cba,Brambilla:2010cs,Cho:1995ce,Cho:1995vh,Baranov:2011ib,Baranov:2016clx,Baranov:2015laa}.
While the production of single quarkonium is understood reasonably
well~\citep{Vanttinen:1998zd,Ivanov:2004vd,Koempel:2011rc,Cui:2018jha,Berezhnoy:1997er},
due to the large number of contributing mechanisms the information
about the target which can be extracted from these data is limited.
Furthermore, the corresponding cross sections frequently involve unknown
NRQCD long-distance matrix elements (LDMEs) that must be extracted
from the same processes, thus limiting the predictive power of the
analysis and reducing it to a mere consistency check. For this reason,
since the early days of QCD, theoretical efforts have been dedicated
to the production mechanisms of multiple quarkonia in the final state~\citep{Brodsky:1986ds,Lepage:1980fj,Berger:1986ii,Baek:1994kj},
which can provide much more detailed information about the gluonic
content of the target. Since the cross sections for multiple charmonium
production decrease rapidly with the number of produced charmonia,
current theoretical and experimental efforts focus primarily on the
production of quarkonium pairs~\citep{Lansberg:2019adr,Ko:2010xy,Baranov:2011zz,Likhoded:2015zna,Shao:2016wor}.
Available experimental data~\citep{LHCb:2023qgu,D0:2015dyx} demonstrate
that simple gluon-gluon fusion is insufficient to describe the inclusive
hadroproduction cross section, signaling significant contributions
from poorly known double-parton scattering (DPS) mechanisms. While
DPS could potentially resolve the observed discrepancy, the required
magnitude of DPS varies significantly across different channels: the
value of the effective cross section $\sigma_{\text{eff}}$, which
parameterizes the magnitude of double-parton scattering, varies by
a factor of 2--3. This large uncertainty indicates that other production
mechanisms may play an important role and warrant detailed analysis.

In this paper, we focus on the contribution of the fragmentation mechanism,
which was introduced in the Collins-Soper formalism~\citep{Collins:1981uw}.
Although this mechanism typically dominates at very high energies,
it is frequently overlooked in analyses of quarkonium pair production.
The underlying assumption of the fragmentation picture is that the
formation time of the final-state mesons (the fragmentation time)
significantly exceeds the coherence (Ioffe) time $t_{c}$ of $Q\bar{Q}$
pair formation~\citep{Altinoluk:2014eka,Ioffe:1969kf} as well as
the shockwave interaction time with the target, $t_{h}$. The nonperturbative
dynamics of the final-state hadronization are encoded in universal
(hadron-dependent) fragmentation functions~\citep{Collins:2011zzd,Navas:2024X}.
Since the latter are essentially nonperturbative objects, at present
most of their parametrizations rely on purely phenomenological extractions
and carry sizable uncertainties. The situation is different for heavy
quarkonium because the heavy-quark mass serves as a natural hard scale
and justifies the use of a perturbative framework over a wide kinematic
range~\citep{Dai:2023rvd,vonKuk:2024uxe,Ma:2013yla}. In what follows,
we distinguish between single- and double-hadron (dihadron) fragmentation,
where a heavy quark $Q$ fragments into one or two quarkonium states,
respectively. The latter mechanism should not be confused with the
so-called double-parton fragmentation discussed in Refs.~\citep{Kang:2011mg,Fleming:2012wy,Fleming:2013qu,Ma:2013yla},
which assumes the fusion of a quark-antiquark pair into a single quarkonium
state and thus contributes to the direct production mechanism.

The role of the fragmentation mechanism in inclusive quarkonium production
has been discussed in Refs.~\citep{Ma:2014mri,Baranov:2007dw}, and
in the context of double quarkonium production in Refs.~\citep{Baranov:2022zgg,Prokhorov:2020owf}.
However, those studies focused primarily on the single-fragmentation
mechanism and were performed within the $k_{T}$-factorization framework,
which is valid at relatively large values of $x$. In this manuscript,
we focus on the role of double-quarkonium fragmentation and its contribution
to inclusive quarkonium pair production in high-energy $pp$ collisions
at the LHC. To account for the potential onset of gluon saturation
effects at these energies, we use the Color Glass Condensate (CGC)
framework~\citep{GLR,McLerran:1993ni,McLerran:1993ka,McLerran:1994vd,MUQI,MV,gbw01:1,Kopeliovich:2002yv,Kopeliovich:2001ee,Gelis:2010nm},
which naturally incorporates saturation effects and provides a phenomenologically
successful description of both hadron-hadron and lepton-hadron collisions~\citep{Kovchegov:1999yj,Kovchegov:2006vj,Balitsky:2008zza,Kovchegov:2012mbw,Balitsky:2001re,Cougoulic:2019aja,Aidala:2020mzt}.
Previous studies of quarkonium pairs in the CGC framework have either
focused on direct quarkonium pair production~\citep{Siddikov:2024qvv,Siddikov:2025tzh}
or analyzed pair production combining the CGC with the color evaporation
model~\citep{Fujii:2013gxa}, which can be viewed as a special case
when the fragmentation function is constant.

The paper is structured as follows. In Sec.~\ref{sec:FrameworkDerivation},
we formulate the theoretical framework used to evaluate the contributions
of the fragmentation mechanism. In Sec.~\ref{subsec:Derivation},
we briefly summarize the CGC framework used to describe the interaction
of high-energy partons with the target. In Sec.~\ref{subsec:Fragmentation-function},
we analytically evaluate the dihadron fragmentation functions (DiFFs)
in the heavy-quark mass limit within the NRQCD framework and discuss
their properties (technical details and final expressions are provided
in Appendix~\ref{sec:cutoff-1}). In Sec.~\ref{subsec:CGC-frag},
we demonstrate that within the CGC picture, the cross section of the
fragmentation mechanism can be naturally expressed as a convolution
of the fragmentation functions and dipole production cross sections,
showing that no additional kinematic cuts on transverse momenta are
required for applicability of the heavy-quark fragmentation. In Sec.~\ref{sec:Numer},
we present our numerical estimates for the contributions of the above-mentioned
 fragmentation mechanisms to $S$-wave charmonium pairs and compare
the predicted results with experimental data. Finally, we draw our
conclusions in Sec.~\ref{sec:Conclusions}.

\section{Theoretical framework}

\label{sec:FrameworkDerivation}

High-energy hadronic collisions at the LHC have been used to study
gluon-mediated hadroproduction and photon-mediated production in ultraperipheral
kinematics. In what follows, we focus on the hadroproduction of heavy
quarkonium pairs in forward kinematics, where the process can be described
in the dilute-dense approximation. The cross section can be written
as 
\begin{equation}
\frac{d\sigma_{pp\to\mathcal{Q}_{1}\mathcal{Q}_{2}X}}{dy_{1}d^{2}\boldsymbol{k}^{\perp}_{1}dy_{2}d^{2}\boldsymbol{k}^{\perp}_{2}}\approx\int dx_{1}\,d^{2}\boldsymbol{q}_{\perp}\,\mathcal{F}\,\left(x_{1},\,\boldsymbol{q}_{\perp},\,\mu^{2}\right)\left.\frac{d\sigma\left(g+p\to\mathcal{Q}_{1}+\mathcal{Q}_{2}+X\right)}{dy_{1}d^{2}\boldsymbol{p}^{\perp}_{1}dy_{2}d^{2}\boldsymbol{p}^{\perp}_{2}}\right|_{\boldsymbol{p}^{\perp}_{a}=\boldsymbol{k}^{\perp}_{a}-\beta_{a}\boldsymbol{q}_{\perp}},\label{eq:LTSep-1-1-1}
\end{equation}
where $(y_{a},\boldsymbol{k}^{\perp}_{a})$ are the rapidities and
transverse momenta of the produced quarkonia. The variables $\beta_{a}=p^{+}_{a}/q^{+}$
are the fractions of the incoming gluon's momentum carried by the
quarkonia, and the momenta $\boldsymbol{p}^{\perp}_{a}=\boldsymbol{k}^{\perp}_{a}-\beta_{a}\boldsymbol{q}_{\perp}$
correspond to their transverse momenta with respect to the incoming
gluon. The function $\mathcal{F}(x_{1},\boldsymbol{q}_{\perp},\mu^{2})$
is the transverse-momentum-dependent parton distribution function
(TMD PDF) of the gluon. At very large transverse momenta, we can approximate
$\boldsymbol{p}^{\perp}_{a}=\boldsymbol{k}^{\perp}_{a}-z_{a}\boldsymbol{q}_{\perp}\approx\boldsymbol{k}^{\perp}_{a}$
and reduce Eq.~(\ref{eq:LTSep-1-1-1}) to 
\begin{equation}
\frac{d\sigma_{pp\to\mathcal{Q}_{1}\mathcal{Q}_{2}X}}{dy_{1}d^{2}\boldsymbol{k}^{\perp}_{1}dy_{2}d^{2}\boldsymbol{k}^{\perp}_{2}}\approx\int dx_{1}\,x_{1}g\,\left(x_{1},\,\,\mu^{2}\right)\left.\frac{d\sigma\left(g+p\to\mathcal{Q}_{1}+\mathcal{Q}_{2}+X\right)}{dy_{1}d^{2}\boldsymbol{p}^{\perp}_{1}dy_{2}d^{2}\boldsymbol{p}^{\perp}_{2}}\right|_{\boldsymbol{p}^{\perp}_{a}\approx\boldsymbol{k}^{\perp}_{a}},\label{eq:LTSep-1-1}
\end{equation}
where we introduced the collinear gluon PDF $x_{1}g(x_{1},\mu^{2})=\int d^{2}\boldsymbol{q}_{\perp}\mathcal{F}(x_{1},\boldsymbol{q}_{\perp},\mu^{2})$,
in agreement with the findings of Refs.~\citep{Gelis:2003vh,Fujii:2006ab,Fujii:2020bkl,Iancu:2003uh}.
In essence, the collinear approximation~(\ref{eq:LTSep-1-1}) reflects
the fact that the typical transverse momentum of a gluon inside the
proton is significantly smaller than the invariant mass of the quarkonium
pair. Alternatively, if we integrate both sides of Eq.~(\ref{eq:LTSep-1-1-1})
over $\boldsymbol{k}^{\perp}_{1}$ and $\boldsymbol{k}^{\perp}_{2}$,
and shift the integration variables on the right-hand side as $\boldsymbol{k}^{\perp}_{a}\to\boldsymbol{p}^{\perp}_{a}+z_{a}\boldsymbol{q}_{\perp}$,
we can separate the integrals over $\boldsymbol{q}_{\perp}$, $\boldsymbol{p}^{\perp}_{1}$,
and $\boldsymbol{p}^{\perp}_{2}$ to obtain the transverse-momentum-integrated
analog of Eq.~(\ref{eq:LTSep-1-1-1}): 
\begin{equation}
\frac{d\sigma_{pp\to\mathcal{Q}_{1}\mathcal{Q}_{2}X}}{dy_{1}\,dy_{2}}\approx\int dx_{1}\,\,x_{1}g\left(x_{1},\,\,\mu^{2}\right)\frac{d\sigma\left(g+p\to\mathcal{Q}_{1}+\mathcal{Q}_{2}+X\right)}{dy_{1}dy_{2}}.\label{eq:LTSep-1-1-2}
\end{equation}
As we will see below, for heavy quarks of mass $m_{Q}$ produced in
forward kinematics, the variable $x_{1}$ remains relatively large
($x_{1}\gtrsim10^{-1}$), meaning that the dilute-dense approximation
remains valid even when integrating down to the region of small $k_{\perp}\sim Q_{s}(x)$,
where $Q_{s}(x)\lesssim m_{Q}$ is the saturation scale. For light
quarks with mass $m_{q}\ll Q_{s}(x)$, the relation~(\ref{eq:LTSep-1-1-2})
does not hold, and the collinear approximation is valid only in the
region of large transverse momenta.

To proceed with the evaluations, we  work in the reference frame,
where the colliding protons have the same energies $E_{p}$ but move
in opposite directions along axis $\slashed{z}$. In what follows,
we write out the explicit light-cone decompositions of the participant
momenta.We denote the momenta of the colliding hadrons (the dilute
projectile and the dense target) as $P_{1}$ and $P_{2}$, the momentum
of the gluon emitted from the projectile as $q$, and the four-momenta
of the produced heavy quarkonia as $p_{1}$ and $p_{2}$. The light-cone
expansion of these momenta is given by 
\begin{eqnarray}
q &  & =\,\left(q^{+},\,0,\,\,\boldsymbol{q}_{\perp}\right),\quad q^{+}\approx x_{1}P^{+}_{1}\label{eq:qPhoton}\\
P_{1} &  & =\left(P^{+},\,\frac{m^{2}_{N}}{2P^{+}},\,\,\boldsymbol{0}_{\perp}\right),\quad P_{2}=\left(\frac{m^{2}_{N}}{2P^{+}},\,P^{+},\,\,\boldsymbol{0}_{\perp}\right),\quad P^{+}=\frac{E_{p}+\sqrt{E^{2}_{p}-m^{2}_{N}}}{\sqrt{2}}\approx\sqrt{2}E_{p}\\
p_{a} &  & =\left(\frac{M^{\perp}_{a}\,e^{y_{a}}}{\sqrt{2}}\,,\,\frac{M^{\perp}_{a}e^{-y_{a}}}{\sqrt{2}},\,\,\boldsymbol{p}^{\perp}_{a}\right),\qquad M^{\perp}_{a}\equiv\sqrt{M^{2}_{a}+\left(\boldsymbol{p}^{\perp}_{a}\right)^{2}},\qquad a=1,2,\label{eq:MesonLC}
\end{eqnarray}
where $m_{N}$ is the nucleon mass and $M_{1},M_{2}$ are the masses
of the produced quarkonia. The proton energy $E_{p}$ in symmetric
kinematics is related to the Mandelstam invariant $s=(P_{1}+P_{2})^{2}$
by $E_{p}=\sqrt{s}/2$. In what follows, we are primarily interested
in the high-energy collider kinematics $q^{+},P^{+}\gg\{M_{a},m_{N}\}$,
where the quarkonia are produced with relatively small transverse
momenta. Since the plus-component of the light-cone momentum of any
particle is always positive, for the inclusive production of the quarkonium
pair the variable $q^{+}$ is kinematically bounded by 
\begin{eqnarray}
 &  & q^{+}\gtrsim\frac{M^{\perp}_{1}\,e^{y_{1}}+M^{\perp}_{2}\,e^{y_{2}}}{\sqrt{2}}.\label{eq:qPlus}
\end{eqnarray}
The invariant mass $M_{12}$ of the produced heavy quarkonium pair
is given in terms of these variables by 
\begin{equation}
{M}^{2}_{12}=\left(p_{1}+p_{2}\right)^{2}=M^{2}_{1}+M^{2}_{2}+2\left(M^{\perp}_{1}M^{\perp}_{2}\cosh\Delta y-\boldsymbol{p}^{\perp}_{1}\cdot\boldsymbol{p}^{\perp}_{2}\right),\quad\Delta y=y_{1}-y_{2},\label{eq:M12}
\end{equation}
which, along with the heavy quarkonium masses, serves as a hard scale
justifying the perturbative analysis of the dihadron fragmentation
function. The invariant mass $M_{12}$ also can be used as a convenient
indicator that the produced quarkonia are well separated kinematically.
To suppress the contribution of the near-threshold region in different
observables, which may receive sizable corrections from soft final-state
interactions between the produced quarkonia, we will sometimes present
results by introducing a lower cut on the invariant mass, $M_{12}>M_{1}+M_{2}+\Lambda$
with $\Lambda\approx1$~GeV, or equivalently, on the relative velocity:
\begin{eqnarray}
 &  & v_{{\rm rel}}=\sqrt{1-\frac{p^{2}_{1}p^{2}_{2}}{\left(p_{1}\cdot p_{2}\right)^{2}}}=\sqrt{1-\frac{4M^{2}_{1}M^{2}_{2}}{\left({M}^{2}_{12}-M^{2}_{1}-M^{2}_{2}\right)^{2}}}\gtrsim0.75>2\alpha_{s}\left(m_{c}\right).\label{eq:vrel}
\end{eqnarray}
This condition is automatically satisfied for any transverse momentum
if the rapidity difference between the quarkonia is sufficiently large,
$|\Delta y|>0.8$. The cross section of the process $g+p\to\mathcal{Q}_{1}+\mathcal{Q}_{2}+X$,
which appears on the right-hand side of Eqs.~(\ref{eq:LTSep-1-1-1})--(\ref{eq:LTSep-1-1-2}),
is the central quantity of interest in this study and will be evaluated
using the Color Glass Condensate approach~\citep{GLR,McLerran:1993ni,McLerran:1993ka,McLerran:1994vd,MUQI,MV,gbw01:1,Kopeliovich:2002yv,Kopeliovich:2001ee,Gelis:2010nm}
in Secs.~\ref{subsec:Derivation} and \ref{subsec:CGC-frag}.

\subsection{High-energy scattering in the CGC picture}

\label{subsec:Derivation}

\begin{figure}
\includegraphics[width=5cm]{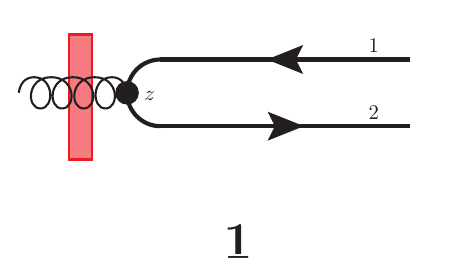}\includegraphics[width=5cm]{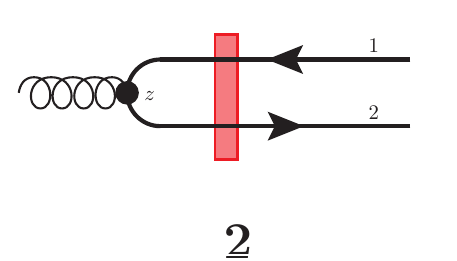}
\caption{ Diagrams contributing to inclusive heavy-quark pair hadroproduction
in the CGC framework at leading order in $\alpha_{s}$. The subscript
$z$ denotes the coordinate of the interaction vertex in configuration
space, and the red block represents the shockwave interaction with
the target.}\label{fig:CGCBasic}
\end{figure}

The interaction of partons with an ultrarelativistic target (shockwave)
in small-$x$ kinematics is characterized by a nearly instantaneous
color exchange. In the eikonal approximation, this can be described
by neglecting the transverse motion of the parton and the change of
its helicity during the interaction. In this picture, the interaction
of heavy partons with the gluonic field of the target is described
by Wilson lines $U(\boldsymbol{x}_{\perp})$ defined as~\citep{GLR,McLerran:1993ni,McLerran:1993ka,McLerran:1994vd,MUQI,MV,gbw01:1,Kopeliovich:2002yv,Kopeliovich:2001ee,Gelis:2010nm}
\begin{equation}
U\left(\boldsymbol{x}_{\perp}\right)=\hat{P}\exp\left(ig\int^{+\infty}_{-\infty}dx^{-}A^{+}_{a}\left(x^{-},\,\boldsymbol{x}_{\perp}\right)t^{a}\right),\label{eq:Wilson}
\end{equation}
where $\boldsymbol{x}_{\perp}$ is the impact parameter of the parton,
$t^{a}$ are the color generators in the corresponding representation
($\boldsymbol{3}$, $\bar{\boldsymbol{3}}$, or $\boldsymbol{8}$
for a quark, antiquark, or gluon, respectively), $A^{+}_{a}(x)$ is
the background gluonic field of the hadron, and operator $\hat{P}$
indicates path-ordering. Formally, this interaction can be represented
via the CGC Feynman rules~\citep{Blaizot:2004,Ayala:2017rmh,Caucal:2021ent,Caucal:2022ulg}
formulated in terms of the effective  vertices ($T$-matrix elements)
describing high-energy parton scattering off the shockwave: 
\begin{equation}
T^{Q}_{\sigma,\sigma';\,i,i'}=2\pi\delta\left(\ell^{+}_{Q}-\ell'{}^{+}_{Q}\right)\gamma^{+}_{\sigma',\,\sigma}\int d^{2}\boldsymbol{z}e^{i\left(\boldsymbol{\ell}_{Q}-\boldsymbol{\ell}_{Q}'\right)\cdot\boldsymbol{z}}\left[U\left(\boldsymbol{z}\right)-1\right]_{i',i}\label{eq:T1}
\end{equation}
\begin{equation}
T^{\bar{Q}}_{\sigma,\sigma';\,i,i'}=-2\pi\delta\left(\ell^{+}_{\bar{Q}}-\ell'{}^{+}_{\bar{Q}}\right)\gamma^{+}_{\sigma,\,\sigma'}\int d^{2}\boldsymbol{z}e^{i\left(\boldsymbol{\ell}_{\bar{Q}}-\boldsymbol{\ell}'_{\bar{Q}}\right)\cdot\boldsymbol{z}}\left[U^{\dagger}\left(\boldsymbol{z}\right)-1\right]_{i,i'}\label{eq:T2}
\end{equation}
\begin{equation}
T^{g}_{\mu,\nu;a,b}=-2\pi\delta\left(\ell^{+}_{g}-\ell'{}^{+}_{g}\right)g_{\mu\nu}{\rm sgn}\left(\ell^{+}_{g}\right)\int d^{2}\boldsymbol{z}e^{-i\left(\boldsymbol{\ell}'_{g}-\boldsymbol{\ell}_{g}\right)\cdot\boldsymbol{z}}\left[\mathcal{U}^{{\rm sgn}\left(\ell^{+}_{g}\right)}\left(\boldsymbol{z}\right)-1\right]_{b,a}
\end{equation}
where $\ell_{i}$ and $\ell'_{i}$ ($i=Q,\bar{Q},g$) are the parton
momenta before and after the interaction, respectively; $\sigma,\sigma'$
are Dirac indices; $i,i'$ are color indices in the fundamental representation
($\boldsymbol{3}$ or $\bar{\boldsymbol{3}}$); and $a,b$ are adjoint
color indices. The matrices $U$ and $\mathcal{U}$ represent the
Wilson lines in the fundamental and adjoint representations, respectively.
The factor $2\pi\delta(\ell^{+}_{i}-\ell'_{i}{}^{+})$ in the vertices
reflects the conservation of longitudinal momentum in the eikonal
approximation. In the classical CGC framework~\citep{McLerran:1993ni,McLerran:1993ka,McLerran:1994vd},
physical observables are averaged over the color charge configurations
$\rho$ in the target with a weight functional $W_{Y}[\rho]$. For
many processes, this averaging can be expressed in terms of universal,
process-independent multipole $S$-matrix elements~\citep{Gelis:2010nm,Kovchegov:2012mbw,Iancu:2003uh}:
\begin{equation}
S_{2n}(Y,\,\boldsymbol{x}_{1},\,\boldsymbol{\xi}_{1},...,\boldsymbol{x}_{n},\,\boldsymbol{\xi}_{n})=\frac{1}{N_{c}}\left\langle {\rm tr}\left(U\left(\boldsymbol{\boldsymbol{x}}_{1}\right)U^{\dagger}\left(\boldsymbol{\xi}_{1}\right)...U\left(\boldsymbol{\boldsymbol{x}}_{n}\right)U^{\dagger}\left(\boldsymbol{\xi}_{n}\right)\right)\right\rangle _{Y},\label{eq:MultipoleDfinition}
\end{equation}
which correspond to dipoles for $n=1$, quadrupoles for $n=2$, etc.
For example, the cross section for heavy-quark pair production can
be obtained by squaring the sum of the diagrams shown in Fig.~\ref{fig:CGCBasic},
projecting the final-state quarks onto states with definite momenta
and averaging over color sources using.~(\ref{eq:MultipoleDfinition}).
The final result of this procedure yields (see Refs.~\citep{Fujii:2020bkl,Ayala:2017rmh,Caucal:2021ent,Caucal:2022ulg}
for details)

\begin{eqnarray}
\frac{d\sigma\left(g+p\to Q+\bar{Q}+X\right)}{dy_{1}d^{2}\boldsymbol{p}^{\perp}_{1}dy_{2}d^{2}\boldsymbol{p}^{\perp}_{2}} &  & =\frac{1}{\left(2\pi\right)^{4}}\int d\alpha\int d^{8}\boldsymbol{X}\,\,\mathcal{R}_{g}\left(\alpha,\boldsymbol{r}_{12}',\,\boldsymbol{r}_{12}\right)\Xi_{{\rm LO}}\left(\boldsymbol{x}_{1},\,\boldsymbol{x}_{2},\,\boldsymbol{x}_{1}',\,\boldsymbol{x}_{2}'\right)\times\label{eq:XSection}\\
 &  & \times e^{-i\left(\boldsymbol{p}_{1\perp}+\boldsymbol{p}_{2\perp}\right)\cdot\left(\boldsymbol{b}_{12}-\boldsymbol{b}_{12}'\right)}e^{-i\left(\bar{\alpha}\boldsymbol{p}_{1\perp}-\alpha\boldsymbol{p}_{2\perp}\right)\cdot\left(\boldsymbol{r}_{12}-\boldsymbol{r}_{12}'\right)},\nonumber 
\end{eqnarray}
where $\alpha$ is the light-cone momentum fraction of the incoming
gluon carried by the quark, $\boldsymbol{x}_{1}$ and $\boldsymbol{x}_{2}$
are the transverse coordinates of the quark and antiquark in the amplitude,
while $\boldsymbol{x}_{1}'$ and $\boldsymbol{x}_{2}'$ are the corresponding
variables in the conjugate amplitude. We define the phase-space volume
element as $d^{8}\boldsymbol{X}=d^{2}\boldsymbol{x}_{1}d^{2}\boldsymbol{x}_{2}d^{2}\boldsymbol{x}_{1}'d^{2}\boldsymbol{x}_{2}'$,
and introduce the dipole sizes and impact parameters: 
\begin{equation}
\boldsymbol{r}_{12}=\boldsymbol{x}_{1}-\boldsymbol{x}_{2},\quad\boldsymbol{r}_{12}'=\boldsymbol{x}_{1}'-\boldsymbol{x}_{2}',\qquad\boldsymbol{b}_{12}=\alpha\boldsymbol{x}_{1}+\bar{\alpha}\boldsymbol{x}_{2},\quad\boldsymbol{b}_{12}'=\alpha\boldsymbol{x}_{1}'+\bar{\alpha}\boldsymbol{x}_{2}',\label{eq:defVars}
\end{equation}
with $\bar{\alpha}\equiv1-\alpha$. The quantity $\mathcal{R}_{g}$
is the perturbative impact factor, given explicitly by 
\begin{align}
\mathcal{R}_{g}\left(\alpha,\boldsymbol{r}_{12}',\,\boldsymbol{r}_{12}\right) & =\frac{\alpha_{s}}{4\pi^{2}}\alpha(1-\alpha)\left\{ m^{2}\,K_{1}\left(mr_{12}\right)K_{1}\left(mr_{12}'\right)\left[e^{i\theta_{12}}\,\alpha^{2}+e^{-i\theta_{12}}(1-\alpha)^{2}\right]\right.\\
 & \left.+m^{2}K_{0}\left(mr_{12}\right)K_{0}\left(mr_{12}'\right)\right\} ,\nonumber 
\end{align}
where $m$ is the heavy-quark mass, $\alpha_{s}$ is the running coupling,
and $\theta_{12}$ is the angle between the transverse vectors $\boldsymbol{r}_{12}$
and $\boldsymbol{r}_{12}'$. The physically relevant real part of
the expression $\left[e^{i\theta_{12}}\,\alpha^{2}+e^{-i\theta_{12}}(1-\alpha)^{2}\right]$
can be rewritten as $\left[\alpha^{2}+(1-\alpha)^{2}\right]\cos\theta_{12}=\left[\alpha^{2}+(1-\alpha)^{2}\right]\hat{\boldsymbol{r}}_{12}\cdot\hat{\boldsymbol{r}}_{12}'$,
in agreement with Refs.~\citep{Caucal:2021ent,Caucal:2022ulg}. The
impact factor $\mathcal{R}_{g}$ can also be represented as a convolution
of the gluon light-cone wave functions. The function $\Xi_{\text{LO}}$
encodes the interaction of the partons with the target and is given
by~\citep{Caucal:2021ent,Caucal:2022ulg,Fujii:2020bkl} 
\begin{align}
\Xi_{{\rm LO}} & =\frac{1}{N^{2}_{c}-1}\left[\frac{}{}-S_{4}\left(Y,\boldsymbol{x}_{1},\,\boldsymbol{x}_{2},\,\boldsymbol{x}_{2}',\,\boldsymbol{x}_{1}'\right)+N^{2}_{c}S_{2}\left(Y,\boldsymbol{x}_{2}',\boldsymbol{x}_{2}\right)S_{2}\left(Y,\boldsymbol{x}_{1},\,\boldsymbol{x}_{1}'\right)-N^{2}_{c}S_{2}\left(Y,\boldsymbol{x}_{1},\,\boldsymbol{b}_{12}'\right)S_{2}\left(Y,\boldsymbol{b}_{12}',\,\boldsymbol{x}_{2}\right)\right.\\
 & \qquad-N^{2}_{c}S_{2}\left(Y,\boldsymbol{x}_{1}',\,\boldsymbol{b}_{12}\right)S_{2}\left(Y,\boldsymbol{b}_{12},\,\boldsymbol{x}_{2}'\right)+S_{2}\left(Y,\boldsymbol{x}_{1},\,\boldsymbol{x}_{2}\right)+S_{2}\left(Y,\boldsymbol{x}_{1}',\,\boldsymbol{x}_{2}'\right)\nonumber \\
 & \qquad\left.+N^{2}_{c}S_{2}\left(Y,\boldsymbol{b}_{12},\,\boldsymbol{b}_{12}'\right)S_{2}\left(Y,\boldsymbol{b}_{12}',\,\boldsymbol{b}_{12}\right)-1\frac{}{}\right],\nonumber 
\end{align}
where we have taken into account that, in the eikonal picture, the
transverse coordinate of the incoming gluon coincides with the impact
parameter of the produced dipole~\citep{Bjorken:1970ah}. Consequently,
terms containing $\boldsymbol{b}_{12}$ or $\boldsymbol{b}_{12}'$
represent the interaction of the shockwave with the incoming gluon
in the amplitude and the conjugate amplitude, respectively. The dipole
$S$-matrix $S_{2}(\boldsymbol{x}_{q},\boldsymbol{x}_{\bar{q}})$
is typically expressed in terms of the forward dipole scattering amplitude
$N(Y,\boldsymbol{r},\boldsymbol{b})$ as 
\begin{equation}
N\left(Y,\,\boldsymbol{r},\,\boldsymbol{b}\right)=1-S_{2}\left(Y,\,\boldsymbol{x}_{q},\,\boldsymbol{x}_{\bar{q}}\right),\label{eq:NS}
\end{equation}
where $\boldsymbol{r}\equiv\boldsymbol{x}_{q}-\boldsymbol{x}_{\bar{q}}$
is the transverse size of the dipole, and $\boldsymbol{b}\equiv\alpha_{q}\,\boldsymbol{x}_{q}+\alpha_{\bar{q}}\boldsymbol{x}_{\bar{q}}$
is the impact parameter of its center of mass~\citep{Bjorken:1970ah}.
The quadrupole contribution $S_{4}\left(Y,\,\boldsymbol{x}_{1},\,\boldsymbol{x}_{2},\,\boldsymbol{x}_{2}',\,\boldsymbol{x}_{1}'\right)$
is much less understood and phenomenologically constrained (see, however,
model-dependent parameterizations in Refs.~\citep{Iancu:2011nj,Iancu:2011ns,Iancu:2012xa}).
Fortunately, for the $\boldsymbol{p}_{\perp}$-integrated cross section,
this quadrupole uncertainty is irrelevant because the quadrupole contribution
reduces to a dipole, namely 
\begin{eqnarray}
\frac{d\sigma\left(g+p\to Q+\bar{Q}+X\right)}{dy_{1}dy_{2}} &  & =\int d^{4}\boldsymbol{X}\,\mathcal{R}_{g}\left(\alpha,\boldsymbol{r}_{12},\,\boldsymbol{r}_{12}\right)\overline{\Xi_{{\rm LO}}\left(\boldsymbol{x}_{1},\,\boldsymbol{x}_{2}\right)},\label{eq:XSecInt}\\
d^{4}\boldsymbol{X} &  & =d^{2}\boldsymbol{x}_{1}d^{2}\boldsymbol{x}_{2}=d^{2}\boldsymbol{b}_{12}\,d^{2}\boldsymbol{r}_{12},\nonumber 
\end{eqnarray}
where 
\begin{align}
\overline{\Xi_{{\rm LO}}} & =-\frac{2}{N^{2}_{c}-1}\left[\left(1-S_{2}\left(Y,\boldsymbol{x}_{1},\,\boldsymbol{x}_{2}\right)\right)+N^{2}_{c}\left[S_{2}\left(Y,\boldsymbol{x}_{1},\,\boldsymbol{b}_{12}\right)S_{2}\left(Y,\,\boldsymbol{b}_{12},\,\boldsymbol{x}_{2}\right)-1\right]\right]=\\
 & -\frac{2N^{2}_{c}}{N^{2}_{c}-1}\left[\frac{1}{N^{2}_{c}}N\left(Y,\boldsymbol{r}_{12},\,\boldsymbol{b}_{12}\right)+N\left(Y,\bar{\alpha}\,\boldsymbol{r}_{12},\,\boldsymbol{b}_{1,12}\right)N\left(Y,-\alpha\boldsymbol{r}_{12},\,\boldsymbol{b}_{2,12}\right)\right.\nonumber \\
 & \qquad-\left.N\left(Y,\bar{\alpha}\,\boldsymbol{r}_{12},\,\boldsymbol{b}_{1,12}\right)-N\left(Y,-\alpha\boldsymbol{r}_{12},\,\boldsymbol{b}_{2,12}\right)\frac{}{}\right]\nonumber 
\end{align}
and in the last line we simplified the arguments of the dipole amplitude
$N$ using a set of identities 
\begin{equation}
\boldsymbol{x}_{1}-\boldsymbol{b}_{12}=\bar{\alpha}\left(\boldsymbol{x}_{1}-\boldsymbol{x}_{2}\right),\quad\boldsymbol{x}_{2}-\boldsymbol{b}_{12}=-\alpha\left(\boldsymbol{x}_{1}-\boldsymbol{x}_{2}\right),
\end{equation}
\begin{equation}
\boldsymbol{b}_{1,12}=\frac{\alpha\boldsymbol{x}_{1}+\boldsymbol{b}_{12}}{1+\alpha}=\boldsymbol{b}_{12}+\frac{\alpha\bar{\alpha}\boldsymbol{r}_{12}}{1+\alpha},
\end{equation}
\begin{equation}
\boldsymbol{b}_{2,12}=\frac{\bar{\alpha}\boldsymbol{x}_{2}+\boldsymbol{b}_{12}}{1+\bar{\alpha}}=\boldsymbol{b}_{12}-\frac{\alpha\bar{\alpha}\boldsymbol{r}_{12}}{1+\bar{\alpha}}.
\end{equation}
To avoid the uncertainty associated with the quadrupole parameterization,
we focus on the $\boldsymbol{p}_{\perp}$-integrated cross sections
in the remainder of this work.

\subsection{Fragmentation function of the quarkonium}

\label{subsec:Fragmentation-function}

The concept of fragmentation functions was introduced in the Collins-Soper
formalism~\citep{Collins:1981uw}. These functions are essentially
nonperturbative objects. Although they are generally frame-independent,
a suitable choice of reference frame simplifies their formal definitions.
For the single-quarkonium collinear fragmentation function, it is
convenient to choose a frame where the final-state quarkonium has
no transverse momentum, $\boldsymbol{p}^{\perp}_{\mathcal{Q}}=\boldsymbol{0}_{\perp}$.
In this frame, the leading-twist fragmentation function $D^{\mathcal{Q}/q}_{1}(z,\mu)$
is defined as~\citep{Metz:2016swz,Navas:2024X,Collins:2011zzd} 
\begin{align}
D^{\mathcal{Q}/q}_{1}(z,\,\mu) & =\frac{z}{4}\sum_{X}\hspace{-0.5cm}\int\;\int\frac{d\xi^{-}}{2\pi}\,e^{ik^{+}\xi^{-}}\textrm{Tr}\,\Big[\,\langle0|\,{\cal W}(\infty^{-},\xi^{-})\,\psi_{q}(0^{+},\xi^{-},\boldsymbol{0}_{T})\,|p_{\mathcal{Q}};\,X\rangle\label{e:corr_quark_kT}\\
 & \times\;\langle p_{\mathcal{Q}};X|\,\bar{\psi}_{q}(0^{+},0^{-},\boldsymbol{0}_{T})\,{\cal W}(0^{-},\infty^{-})\,|0\rangle\gamma^{-}\Big]\,,\nonumber 
\end{align}
where $\psi_{q}$ and $\bar{\psi}_{q}$ are the quark and antiquark
field operators, and $\mathcal{W}$ are the usual gauge links defined
as 
\begin{equation}
{\cal W}(a^{-},b^{-})=\hat{P}\,\textrm{exp}\bigg[ig\int^{\left(0,\,b,\,\boldsymbol{0}_{\perp}\right)}_{\left(0,\,a,\,\boldsymbol{0}_{\perp}\right)}d\zeta\,A^{+}_{a}(\zeta)t^{a}\bigg].\label{eq:gauge_link}
\end{equation}
The variable $z=p^{+}_{\mathcal{Q}}/k^{+}$ is the light-cone momentum
fraction of the parent quark carried by the quarkonium $\mathcal{Q}$.
The dependence of the fragmentation function on the factorization
scale $\mu$ is governed by the DGLAP evolution equations. In this
reference frame, the transverse-momentum-dependent fragmentation function
(TMD FF) $D^{\mathcal{Q}/q}_{1}(z,\boldsymbol{k}_{\perp},\mu)$ is
a straightforward generalization with a nonzero transverse separation
and a subsequent Fourier transform over it, where $\boldsymbol{k}_{\perp}$
represents the transverse momentum of the quark prior to fragmentation.

For the unpolarized dihadron (diquarkonium) fragmentation function,
the definition is simplest in the reference frame where the total
momentum of the quarkonium pair $P^{\mu}_{h}=p^{\mu}_{1}+p^{\mu}_{2}$
does not have transverse component, namely $\boldsymbol{p}_{1\perp}+\boldsymbol{p}_{2\perp}=\boldsymbol{0}_{\perp}$,
and the light-cone decomposition of the four-momenta of quarkonia
is given by 
\begin{equation}
p_{a}=\left(\frac{\left(1\pm\zeta\right)}{2}P^{+}_{h},\quad\frac{M^{2}_{a}+R^{2}_{\perp}}{\left(1\pm\zeta\right)P^{+}_{h}},\quad\pm\boldsymbol{R}_{\perp}\right),\quad a=1,2;\quad\zeta=\frac{p^{+}_{1}-p^{+}_{2}}{p^{+}_{1}+p^{+}_{2}}.
\end{equation}
 The invariant mass of the produced quarkonia system in this frame
is given by 
\begin{align}
M^{2}_{12} & =\left(p_{1}+p_{2}\right)^{2}=2\left(\frac{M^{2}_{1}}{1+\zeta}+\frac{M^{2}_{2}}{1-\zeta}\right)+\frac{4R^{2}_{\perp}}{1-\zeta^{2}}\ge2\left(\frac{M^{2}_{1}}{1+\zeta}+\frac{M^{2}_{2}}{1-\zeta}\right),\label{eq:M12-1}
\end{align}
and the dihadron fragmentation function is defined as~\citep{Metz:2016swz}
\begin{align}
D^{\mathcal{Q}_{1}\mathcal{Q}_{2}/q}_{1}\left(z,\zeta\right) & =\frac{z}{4}\,\sum_{X}\hspace{-0.5cm}\int\;\int\frac{d\xi^{-}}{2\pi}\,e^{ik^{+}\xi^{-}}\,\textrm{Tr}\,\Big[\langle0|\,{\cal W}(\infty^{-},\xi^{-})\,\psi_{q}(0^{+},\xi^{-},\boldsymbol{0}_{T})\,|P_{1},P_{2};X\rangle\label{e:D1_quark_DiFF_def-2}\\
 & \times\;\langle P_{1},P_{2};X|\,\bar{\psi}_{q}(0^{+},0^{-},\boldsymbol{0}_{T})\,{\cal W}(0^{-},\infty^{-})\,|0\rangle\,\gamma^{-}\Big]\,,\nonumber 
\end{align}
where $z_{1}$ and $z_{2}$ are the light-cone momentum fractions
carried by the produced quarkonia, with $z=z_{1}+z_{2}$, and 
\begin{equation}
z_{1}=\frac{p^{+}_{1}}{k^{+}}=z\,\frac{1+\zeta}{2}\,,\qquad z_{2}=\frac{p^{+}_{2}}{k^{+}}=z\,\frac{1-\zeta}{2}\,.\label{eq:z1z2Def}
\end{equation}
In general, both single and double fragmentation functions are nonperturbative
and cannot currently be calculated from first principles. However,
for heavy quarkonium, it is possible to justify a perturbative treatment
in the heavy-quark mass limit and evaluate them analytically within
the NRQCD framework. For the single $S$-wave quarkonium production,
the corresponding fragmentation functions have been evaluated in Ref.~\citep{Ma:2013yla}
and are given by
\begin{eqnarray}
D^{\mathcal{Q}/c}_{1}\left(z,\,\mu\right) &  & \approx\alpha^{2}_{s}\left(\mu^{2}\right)\left[\frac{2}{3}\frac{C^{2}_{F}}{N_{c}}\frac{(z-1)^{2}}{(z-2)^{6}}z\left(5z^{4}-32z^{3}+72z^{2}-32z+16\right)\right]\frac{\left\langle \mathcal{\mathcal{O}}_{\mathcal{Q}}\left(^{3}S^{[1]}_{1}\right)\right\rangle }{m^{3}_{c}},\label{eq:Dc}\\
D^{\mathcal{Q}/b}_{1}\left(z,\,\mu\right) &  & \approx\frac{\alpha^{2}_{s}\left(\mu^{2}\right)}{12N_{c}z}\left[\left(z^{2}-2z+2\right)\ln\left[\frac{\mu^{2}_{0}}{4m^{2}_{c}\left(1-z+z^{2}\eta/4\right)}\right]-2z^{2}\left(1+\frac{1-z-z^{2}/2}{4-4z+z^{2}\eta}\eta\right)\right]\frac{\left\langle \mathcal{\mathcal{O}}_{\mathcal{Q}}\left(^{3}S^{[8]}_{1}\right)\right\rangle }{m^{3}_{c}},\label{eq:Db}
\end{eqnarray}
where $\eta=m_{b}/m_{c}$, $\mu_{0}$ is the initial factorization
scale at which the NRQCD matrix elements are defined, and $\langle\mathcal{O}_{\mathcal{Q}}(^{3}S^{[1]}_{1})\rangle$,
$\langle\mathcal{O}_{\mathcal{Q}}(^{3}S^{[8]}_{1})\rangle$ are the
corresponding color-singlet and color-octet LDMEs of the mesons. In
what follows, we disregard the contributions proportional to the color-octet
LDMEs, as they are formally suppressed by $\mathcal{O}(v^{4})$ in
NRQCD~\citep{Cho:1995ce,Cho:1995vh,Bodwin:1994jh} and are numerically
small~\footnote{In potential models, the relative velocity of the quarks in charmonium
is $v\sim\alpha_{s}(m_{c})\approx1/3$. Therefore, the $\mathcal{O}(\alpha_{s})$
expansion of the hard partonic amplitude and the $\mathcal{O}(v)$
expansion of the LDMEs should be considered together when selecting
the dominant contributions, provided the relevant scale in the partonic
amplitude is comparable to $m_{c}$.}. Potentially, the color-octet contributions could be enhanced near
$z\approx0$ due to the $1/z$ pole in Eq.~(\ref{eq:Db}). However,
the residue of this pole is proportional to $\ln(\mu^{2}_{0}/4m^{2}_{c})$,
and for any reasonable choice of the scale $\mu_{0}$ near $2m_{c}$,
the contribution of $D^{\mathcal{Q}/b}_{1}$ remains negligible\footnote{As can be seen from the analytic results in Ref.~\citep{Ma:2013yla},
a similar $1/z$ pole is present in front of all color-octet LDMEs,
but the residue is proportional to $\ln(\mu^{2}/4m^{2}_{c})$, causing
the pole to vanish for $\mu=2m_{c}$.}.

\begin{figure}
\begin{tabular}{ccc}
\includegraphics[width=4.5cm]{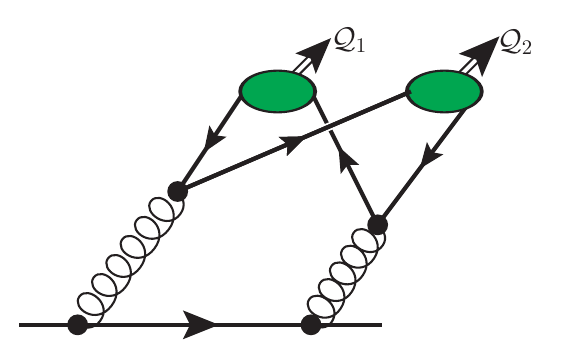}  & \includegraphics[width=5cm]{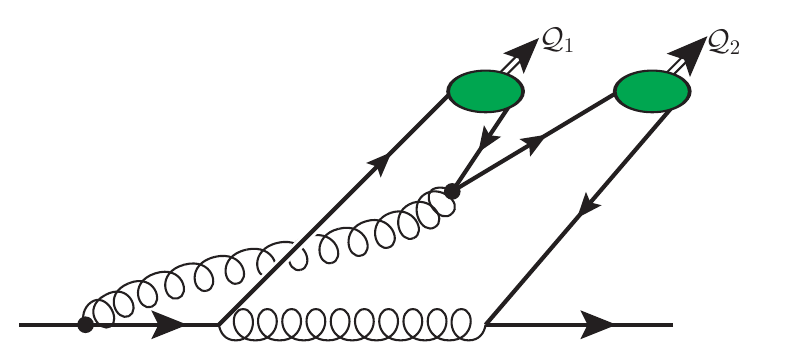}  & \includegraphics[width=5cm]{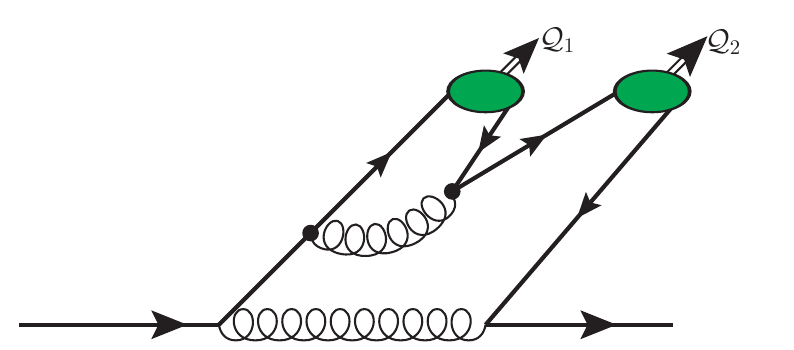}\tabularnewline
\includegraphics[width=5cm]{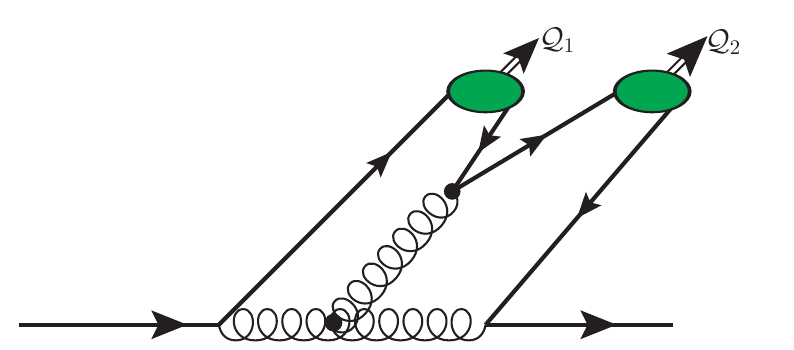}  & \includegraphics[width=5cm]{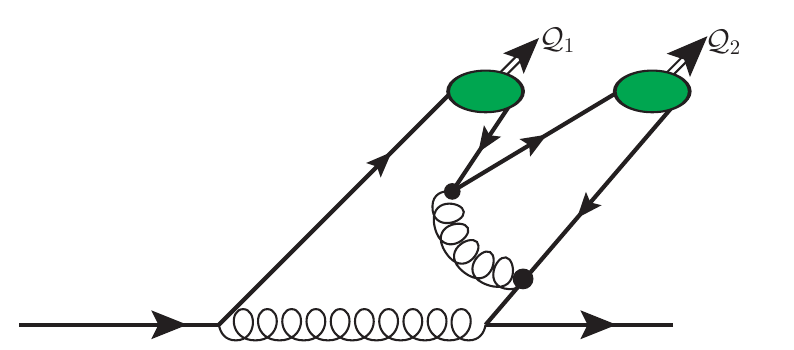}  & \includegraphics[width=5cm]{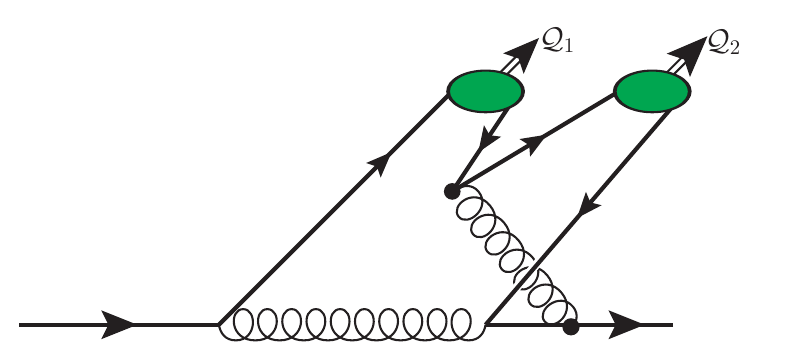}\tabularnewline
\includegraphics[width=5cm]{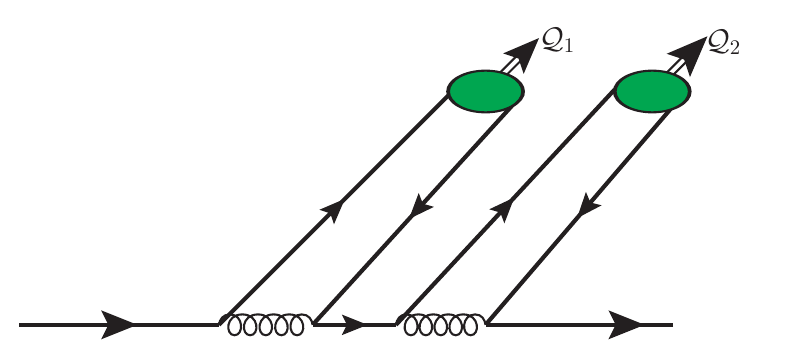}  & \includegraphics[width=5cm]{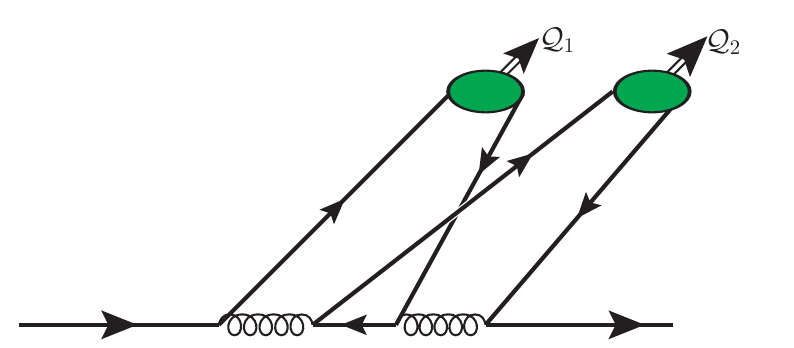}  & \includegraphics[width=5cm]{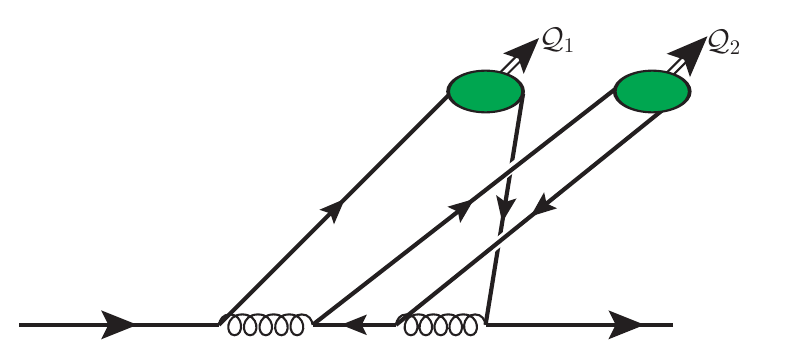}\tabularnewline
\end{tabular}\caption{ Leading-order diagrams contributing to the amplitude of the process
$q\to\mathcal{Q}_{1}\mathcal{Q}_{2}q$ at leading order in $\alpha_{s}$.
Each diagram must be supplemented by a corresponding diagram with
permuted mesons $\mathcal{Q}_{1}\leftrightarrow\mathcal{Q}_{2}$.
If the flavor of the initiating heavy quark does not match the constituent
flavor of the quarkonia (e.g., $b\to J/\psi\,J/\psi\,b$), only the
first diagram contributes. The last two diagrams contribute to $J/\psi$
and $\psi(2S)$ production only via color-octet LDMEs. The full fragmentation
function is proportional to a square of the sum of the corresponding
amplitudes, and may be reduced to a sum of the pairwise products of
different diagrams in the list. }\label{fig:diags_Fragmentation_Charm}
\end{figure}

We can use the same framework to evaluate the fragmentation function
for double-quarkonium production. In what follows, we will use the
light-cone gauge $n\cdot A=A^{+}=0$, so the contributions of the
gauge links~(\ref{eq:gauge_link}) vanish identically. To find the
amplitude of the process $q\to\mathcal{Q}_{1}\mathcal{Q}_{2}q$, we
must take into account all the diagrams shown schematically in Fig.~\ref{fig:diags_Fragmentation_Charm}.
In the heavy-quark mass limit, the quarks carry exactly half of the
quarkonium momentum (neglecting internal motion), so the virtualities
of all intermediate partons are fixed by the external kinematics and
are large ($\sim M^{2}_{12}$), which justifies the perturbative evaluation.
For vector charmonia ($J/\psi,\psi(2S)$), the last two diagrams are
mediated by color-octet LDMEs and are thus negligibly small. The evaluation
is simplest for fragmentation from $b$ quarks, which receives contributions
only from the first diagram in Fig.~\ref{fig:diags_Fragmentation_Charm}.
We perform the evaluation in the heavy-quark mass limit using conventional
NRQCD projectors (see details and the final analytic result in Appendix~\ref{sec:cutoff-1}).
For the fragmentation of the leading charm quark, the evaluation follows
the same lines, but the number of diagrams increases significantly.
A direct evaluation becomes challenging because the full fragmentation
function requires squaring the corresponding amplitude, which involves
a sum of pairwise products of the different diagrams shown in Fig.~\ref{fig:diags_Fragmentation_Charm}.
For this reason, we perform the calculation in the helicity basis,
assuming that for unpolarized quarkonium, the cross section is dominated
by contributions that do not flip the helicity (spin) of the initiating
quark (see details in Appendix~\ref{sec:HelicityBasis}).

\begin{figure}
\includegraphics[width=9cm]{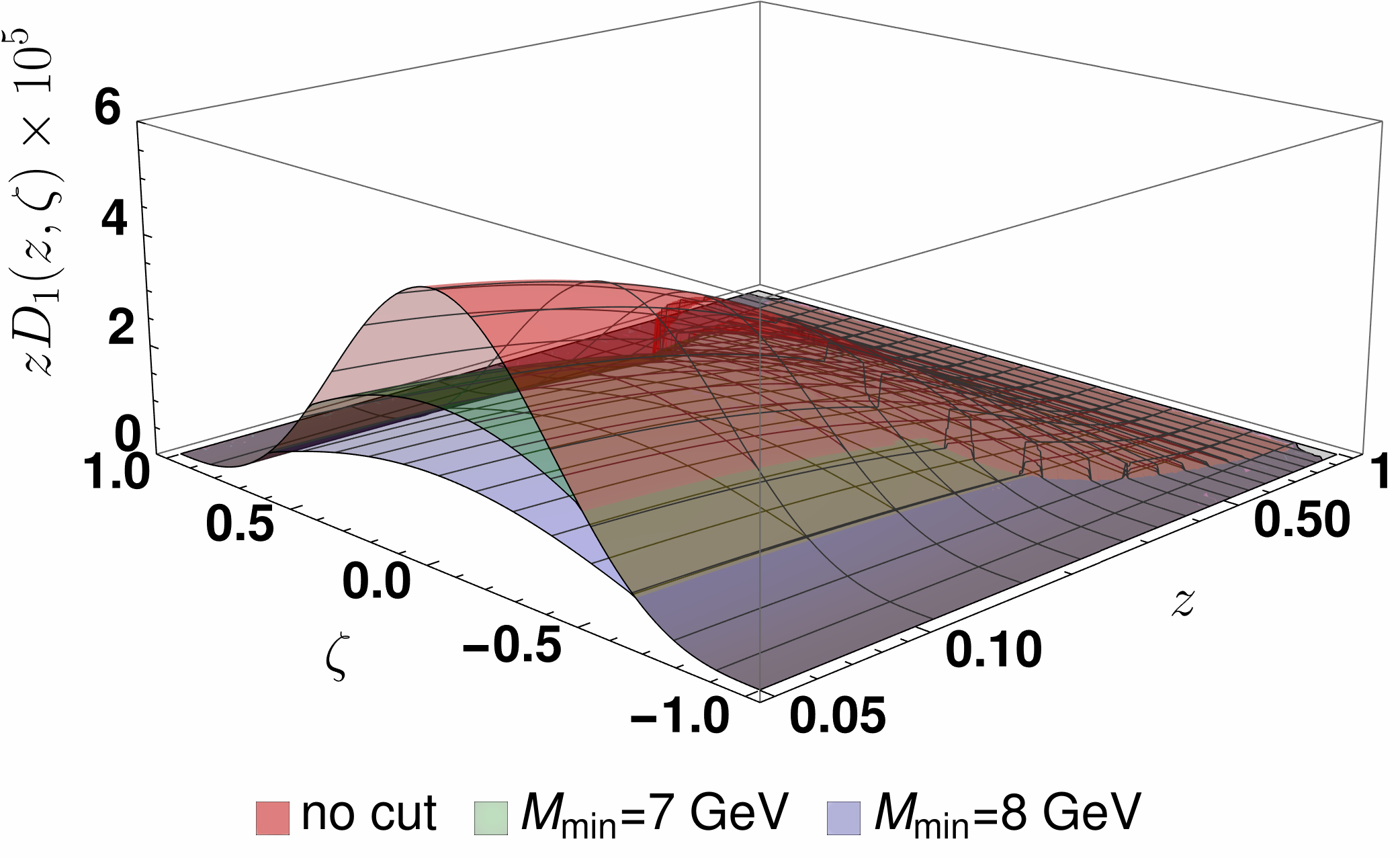}\includegraphics[width=9cm]{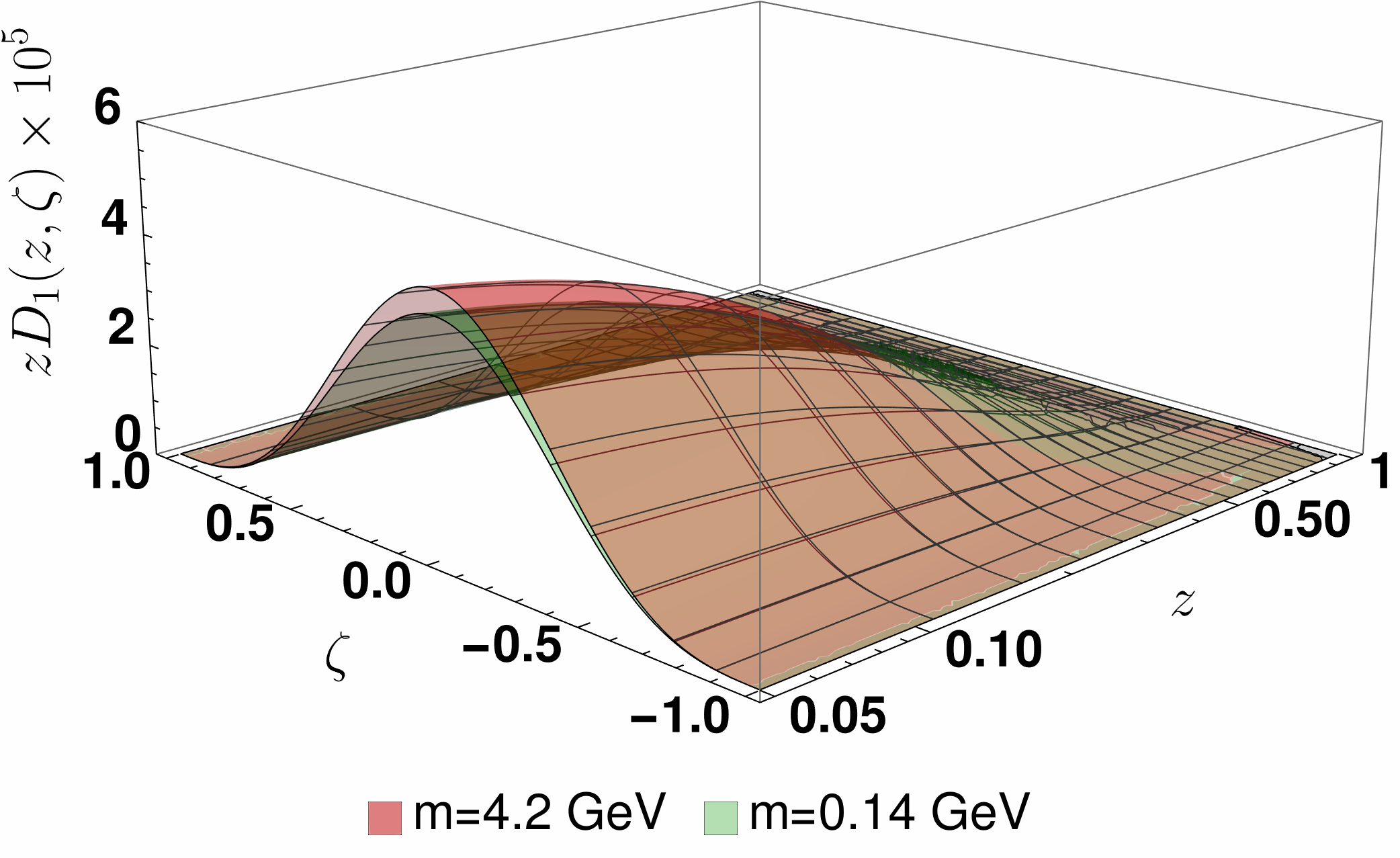}
\includegraphics[width=9cm]{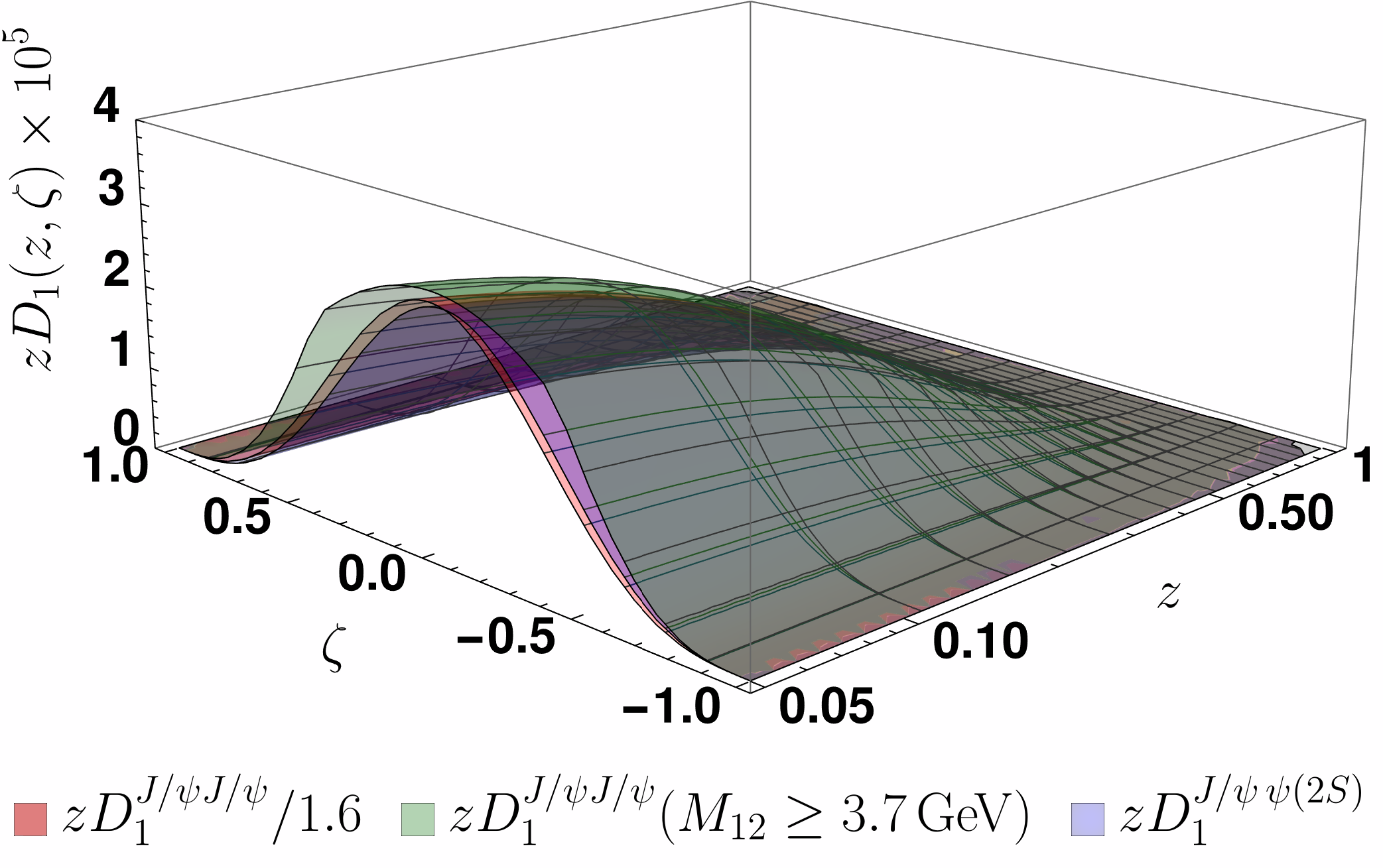}\includegraphics[width=9cm]{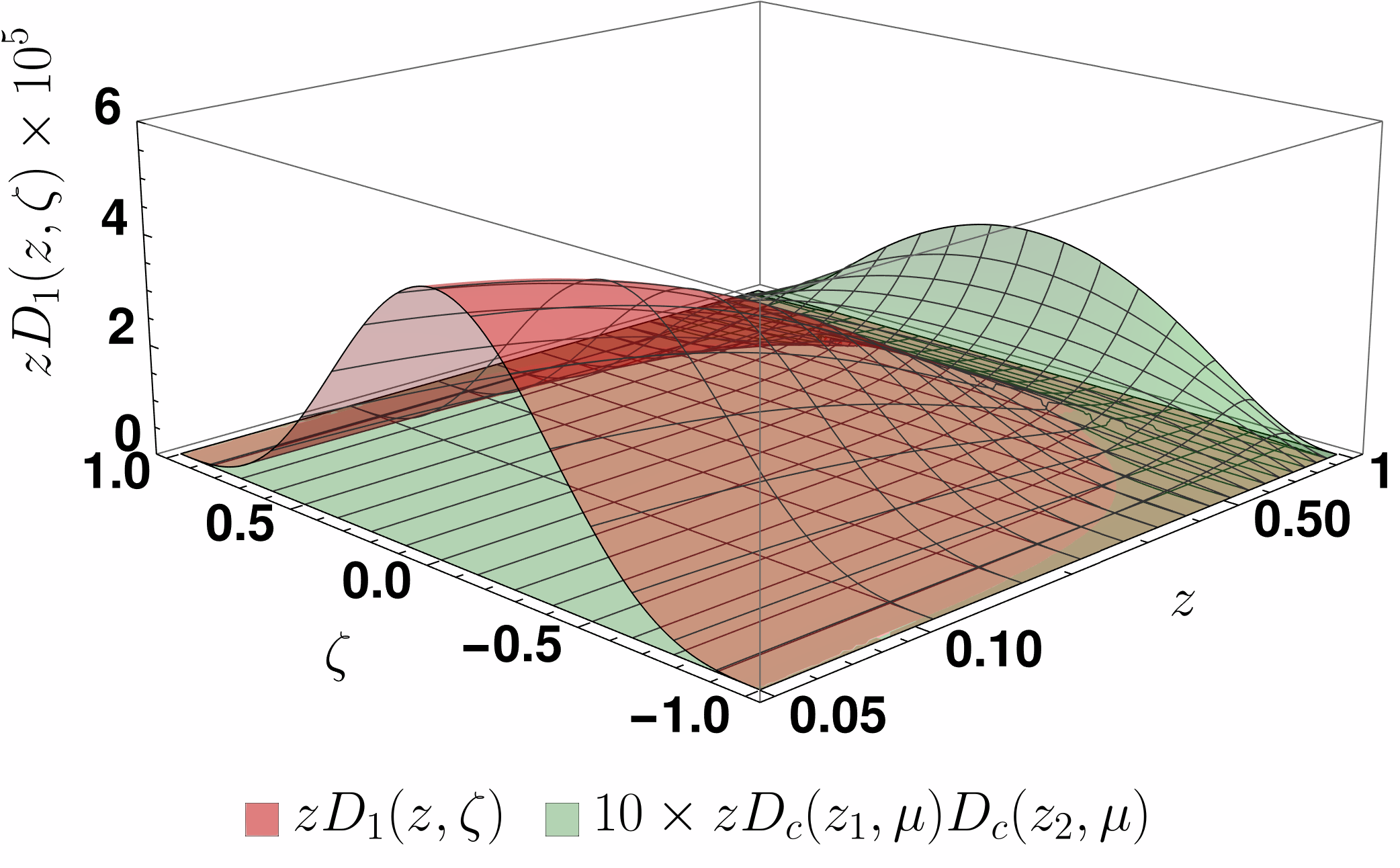}
\caption{ (Color online) Dihadron fragmentation function $zD_{1}(z,\zeta)$
as a function of the variables $(z,\zeta)$. Upper row: Dihadron fragmentation
function for $J/\psi\,J/\psi$ for different choices of the invariant
mass cutoff $M_{\text{min}}$ and the initiating quark mass $m$.
Lower-left plot: Comparison of the dihadron fragmentation functions
for $J/\psi\,J/\psi$ and $J/\psi\,\psi(2S)$. For the $J/\psi\,J/\psi$
case, we show results with an invariant mass cutoff (green) and with
an arbitrary normalization factor $N\approx1/1.6$ to match the normalization
(red). Lower-right plot: Comparison of the dihadron fragmentation
function for $J/\psi\,J/\psi$ with the product of two single-fragmentation
functions $c\to J/\psi$.}\label{fig:diags_Fragmentation_Charm-1}
\end{figure}

The final results of this evaluation are presented in Fig.~\ref{fig:diags_Fragmentation_Charm-1}.
In the upper row, we show the dihadron fragmentation function of $J/\psi\,J/\psi$.
Since our perturbative approach may be unreliable in the near-threshold
region, we analyze the contributions of different kinematic domains
in the upper-left plot by introducing a lower cutoff on the invariant
mass $M_{12}$. While a fragmentation function with a kinematic cutoff
lacks a strict formal definition and violates the principle of universality,
it is useful for understanding the contributions from different regions
and establishing a lower bound on the fragmentation function which
comes from the perturbative domain. We observe that the cutoff suppresses
the dihadron fragmentation function in the central region ($\zeta\approx0$)
but has almost no effect at large $|\zeta|$. Indeed, Eq.~(\ref{eq:M12-1})
shows that for such asymmetric configurations, the invariant mass
is large even for $R_{\perp}=0$, so the cutoff does not alter the
result. The upper-right plot demonstrates that the dependence on the
mass $m$ of the initiating quark is very mild. This is not suprising,
since in most expressions $m$ appears in combination with parametrically
larger variables (of the order of the invariant mass $M^{2}_{12}$).
In the lower-left panel, a comparison of the red and blue surfaces
shows that unequal quarkonium masses introduce a slight asymmetry
in the $\zeta$ dependence of the dihadron fragmentation function.
Finally, in the lower-right panel, we compare the dihadron fragmentation
function with the product of two single-fragmentation functions. The
latter is more than an order of magnitude smaller because single fragmentation
from a bottom quark is governed by the small color-octet LDME, whereas
single fragmentation from a charm quark requires the co-production
of an additional undetected $c$ or $\bar{c}$ quark for each fragmentation
function~\citep{Ma:2013yla}, leading to strong phase-space suppression.

\subsection{Quarkonia production via the fragmentation mechanism in the CGC}

\label{subsec:CGC-frag}

\begin{figure}
\includegraphics[width=7cm]{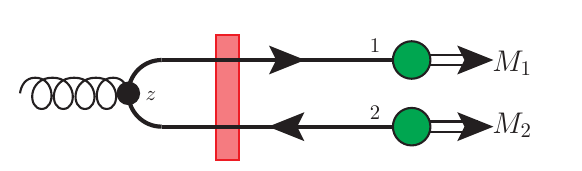}\includegraphics[width=7cm]{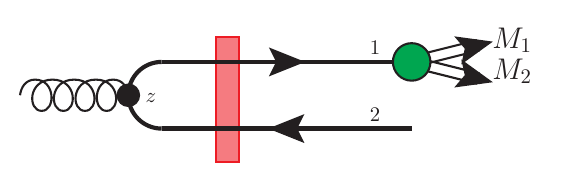} 

\includegraphics[width=9cm]{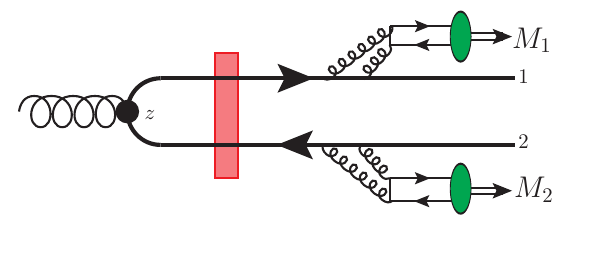}\includegraphics[width=9cm]{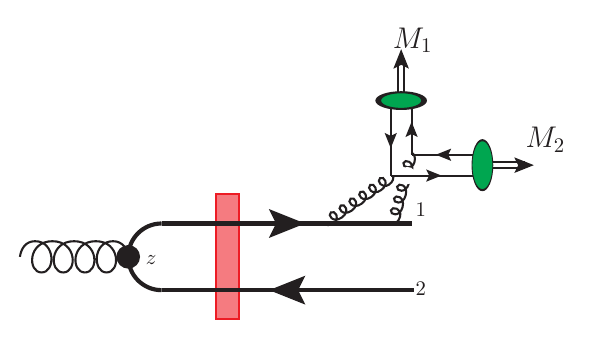}
\caption{ Upper row: Representative diagrams illustrating the single- and
double-fragmentation contributions to inclusive heavy-quark pair hadroproduction
in the CGC framework at leading order in $\alpha_{s}$ (left and right
panels, respectively). Lower row: Representative diagrams corresponding
to single- and dihadron fragmentation. For the quarkonium states considered
here, both the quark and antiquark produced after the shockwave contribute
equally to the fragmentation. The red block represents the shockwave,
which can interact either with the quarks or the incoming gluon.}\label{fig:CGCBasic-1}
\end{figure}

In this section, we show in detail that fragmentation in the CGC framework
corresponds to a specific class of diagrams, shown schematically in
Fig.~\ref{fig:CGCBasic-1}. The main goal of this section is to demonstrate
that the off-shellness of the heavy quarks does not prevent the representation
of the cross section as a convolution of the dipole cross section~(\ref{eq:XSecInt})
with the fragmentation functions~(\ref{e:corr_quark_kT}) and (\ref{e:D1_quark_DiFF_def-2}).
Furthermore, in the heavy-quark mass limit, there is no need to introduce
an artificial lower cutoff on the transverse momenta of the quarks.
However, this analysis does not apply to light quarks, since a large
invariant mass of the quarkonium pair alone is insufficient to justify
a perturbative treatment. For this reason, we consider only the contributions
from heavy flavors ($c$ and $b$), assuming that such events can
be experimentally separated from the light-quark background.

For simplicity, in what follows we consider only the unpolarized (spin-averaged)
production cross sections. Technically, this factorization in the
CGC picture is straightforward to justify and closely follows the
derivation of the $Q\bar{Q}$ cross section in Refs.~\citep{Caucal:2021ent,Caucal:2022ulg,Fujii:2020bkl}.
For definiteness, we demonstrate how factorization occurs for one
specific diagram which corresponds to dihadron fragmentation, assuming
that the proof can be extended to all other contributions. For convenience,
we perform the analysis in the reference frame where the incoming
gluon $q$ is on-shell and moves in the plus direction, $q^{\mu}=(q^{+},0,\boldsymbol{0}_{\perp})$,
assuming that all momenta in Eqs.~(\ref{eq:qPhoton})--(\ref{eq:MesonLC})
can be transformed to this frame by a transverse boost $\boldsymbol{k}_{a\perp}\to\boldsymbol{k}_{a\perp}-\mathfrak{z}_{a}\boldsymbol{q}_{\perp}$,
where $\mathfrak{z}_{a}=k^{+}_{a}/q^{+}$ is the light-cone momentum
fraction of the corresponding particle (which is invariant under such
boosts). The virtual $b$ quark subsequently fragments in the vacuum
into a final-state $b$ quark with momentum $k_{1}$ and two heavy
charmonia with momenta $p_{1}$ and $p_{2}$. Momentum conservation
allows us to reconstruct the momentum of the quark prior to fragmentation
(immediately after its interaction with the shockwave): 
\begin{equation}
K=k_{1}+p_{1}+p_{2}.\label{eq:kRel}
\end{equation}
In the color-singlet NRQCD framework, we neglect the internal relative
motion of the constituent quarks, so the momentum of each charmonium
is shared equally between its constituent $c$ and $\bar{c}$ quarks.
The momenta of the gluons connecting the charm and bottom lines can
then be determined from momentum conservation as 
\begin{equation}
q_{1}=q_{2}=\frac{p_{1}+p_{2}}{2},\qquad q^{2}_{1}=q^{2}_{2}=\frac{M^{2}_{12}}{4}.
\end{equation}
Using the CGC Feynman rules, the cross-section of the quarkonium pair
production can be represented as 
\begin{equation}
\frac{d\sigma\left(g+p\to\mathcal{Q}_{1}+\mathcal{Q}_{2}+X\right)}{dy_{1}dy_{2}}=\frac{1}{16\pi}\int^{1}_{0}d\alpha\int\prod^{2}_{j=1}\frac{d^{2}\boldsymbol{p}_{j}}{2\left(2\pi\right)^{3}}\frac{d^{2}\boldsymbol{k}_{j}}{2\left(2\pi\right)^{3}}\left\langle \left|\mathcal{A}^{a}\right|^{2}\right\rangle _{Y},
\end{equation}
where $\alpha=K^{+}/q^{+}$ , and the variables $k_{1}$ and $k_{2}$
are the momenta of active quark (after fragmentation) and the spectator
antiquark, respectively. The bracket $\left\langle ...\right\rangle _{Y}$
stands for the averaging over the color sources discussed in Section~\ref{subsec:Derivation}.
The notation $\mathcal{A}^{a}$ is used for the amplitude of the $g\to\mathcal{Q}_{1}\mathcal{Q}_{2}b\bar{b}$
subprocess in the field of the shockwave. The latter amplitude can
be represented as
\begin{align}
\mathcal{A}^{a} & =\sqrt{4\pi\alpha_{s}}\int\frac{d^{4}k}{\left(2\pi\right)^{4}}\int d^{2}\mathbf{x}_{1}d^{2}\mathbf{x}_{2}\,e^{i\left(\mathbf{k}{}_{\perp}-\mathbf{K}{}_{\perp}\right)\cdot\mathbf{x}_{1}-i\left(\mathbf{k}_{2\perp}-\mathbf{k}_{\perp}\right)\cdot\mathbf{x}_{2}}\,\,\left[U(\mathbf{x}_{1})t^{a}U^{\dagger}(\mathbf{x}_{2})\,-\mathcal{U}_{ab}\left(\boldsymbol{b}_{12}\right)t^{b}\right]_{\bar{c}c'}\times\label{eq:M}\\
 & \times(2\pi)\delta\left(k^{+}-K^{+}\right)\,\,\bar{u}\left(k_{1}\right)\mathcal{M}^{c,c'}_{\text{frag}}\left(k_{1},p_{1},\,p_{2};K\right)S_{F}\left(K\right)\gamma^{+}S_{F}\left(k\right)\slashed{\varepsilon}\left(q\right)S_{F}\left(q-k\right)\gamma^{+}v\left(k_{2}\right)\nonumber 
\end{align}
where $\varepsilon^{(\lambda)}_{\mu}(q)$ is the polarization vector
of the incoming (primordial) gluon with polarization $\lambda$, and
$\slashed{\varepsilon}\equiv\varepsilon^{(\lambda)}_{\mu}\gamma^{\mu}$.
The contribution $\mathcal{U}_{ab}\left(\boldsymbol{b}_{12}\right)$
corresponds to the interactions of the incoming gluon with transverse
coordinate $\boldsymbol{b}_{12}$ defined in~(\ref{eq:defVars}).
The notation $\mathcal{M}^{c',c}_{\text{frag}}(k_{1},p_{1},p_{2};K)$
stands for the (cut) amplitude of the $b^{(c')}\to b^{(c)}\mathcal{Q}_{1}\mathcal{Q}_{2}$
process (the sum of all the appropriate contributions shown in the
Figure~\ref{fig:diags_Fragmentation_Charm-1} is implied), $c'$
and $c$ are color indices of quark before and after fragmentation~\footnote{For the dominant color singlet LDMEs, the amplitude $\mathcal{M}^{c',c}_{\text{frag}}$
also belongs to the color singlet irreducible representations (namely,
is diagonal w.r.t. color singlet indices). However, for color octet
LDMEs the amplitude may include nondiagonal contributions $\sim\left(t^{a}\right)_{c',c}$.}, and 
\begin{equation}
S_{F}(\ell)=i\frac{\slashed{\ell}+m}{\ell^{2}-m^{2}},\label{eq:freeProp}
\end{equation}
is the free Dirac propagator. It is convenient to rewrite the quark
propagators connected to the shockwave (which are multiplied by $\gamma^{+}$)
as 
\begin{align}
S(\ell) & =\frac{i\left(\slashed{\ell}+m\right)}{\ell^{2}-m^{2}+i0}=\theta\left(\ell^{+}\right)\frac{i\left(\slashed{\ell}'_{Q}+m\right)}{\ell^{2}-m^{2}+i0}-\theta\left(-\ell^{+}\right)\frac{i\left(\slashed{\ell}'_{\bar{Q}}-m\right)}{\ell^{2}-m^{2}+i0}+\frac{\left(\ell^{-}-\frac{\ell^{2}_{\perp}+m^{2}}{2\ell^{+}}\right)\gamma^{+}}{\ell^{2}-m^{2}+i0}=\label{eq:Prop}\\
 & =\frac{\theta\left(\ell^{+}\right)\sum_{h}u_{h}\left(\ell'_{Q}\right)\bar{u}_{h}\left(\ell'_{Q}\right)-\theta\left(-\ell^{+}\right)\sum_{h}v_{h}\left(\ell'_{\bar{Q}}\right)\bar{v}_{h}\left(\ell'_{\bar{Q}}\right)}{\ell^{2}-m^{2}+i0}+\frac{\gamma^{+}}{2\ell^{+}},\nonumber 
\end{align}
where the prime ($'$) denotes that the minus-component of the vector
is adjusted to satisfy the on-shell condition $\ell^{2}=m^{2}$, namely
\begin{equation}
\ell'=\left(\ell^{+},\,\frac{\ell^{2}_{\perp}+m^{2}}{2\ell^{+}},\,\boldsymbol{\ell}_{\perp}\right),\qquad\ell_{Q}\equiv\ell',\qquad\ell'_{\bar{Q}}=-\ell'.
\end{equation}

The last term in Eq.~(\ref{eq:Prop}) corresponds to the instantaneous
propagator in light-cone perturbation theory~\citep{Lepage:1980fj}.
Since the shockwave interaction vertices~(\ref{eq:T1})--(\ref{eq:T2})
contain the matrix $\gamma^{+}$, the identity $\gamma^{+}\gamma^{+}=0$
allows us to neglect the instantaneous term in the propagators connected
to the shockwave. The product of the quark propagators connected to
the shockwave can then be simplified using the identities~\citep{Lepage:1980fj}
\begin{equation}
\bar{u}_{h_{1}}\left(\ell_{1}\right)\gamma^{+}u_{h_{2}}\left(\ell_{2}\right)=\bar{v}_{h_{1}}\left(\ell_{1}\right)\gamma^{+}v_{h_{2}}\left(\ell_{2}\right)=2\sqrt{\ell^{+}_{1}\ell^{+}_{2}}\delta_{h_{1},h_{2}},\label{eq:Id}
\end{equation}
\begin{equation}
S\left(\ell_{1}\right)\gamma^{+}S\left(\ell_{2}\right)=2\sqrt{\ell^{+}_{1}\ell^{+}_{2}}\frac{\theta\left(\ell^{+}_{1}\right)\sum_{h}u_{h}\left(\ell_{1}'\right)\bar{u}_{h}\left(\ell_{2}'\right)-\theta\left(-\ell^{+}_{1}\right)\sum_{h}v_{h}\left(-\ell_{1}'\right)\bar{v}_{h}\left(-\ell_{2}'\right)}{\left(\ell^{2}_{1}-m^{2}+i0\right)\left(\ell^{2}_{2}-m^{2}+i0\right)}.\label{eq:S}
\end{equation}
The dependence on $k^{-}$ in the integrand of Eq.~(\ref{eq:M})
is isolated in the product $S_{F}(k)\slashed{\varepsilon}(q)S_{F}(q-k)$.
Integration over this variable is straightforward using the Cauchy's
residue theorem: 
\begin{align}
 & \int^{+\infty}_{-\infty}\frac{dk^{-}}{2\pi}\frac{1}{\left(2k^{+}k^{-}-\boldsymbol{k}{}^{2}_{\perp}-m^{2}+i0\right)\left(2\left(q^{+}-k^{+}\right)\left(q^{-}-k^{-}\right)-\left(\boldsymbol{q}_{\perp}-\boldsymbol{k}{}_{\perp}\right)^{2}-m^{2}+i0\right)}=\label{eq:Cauchy_2}\\
 & =-\frac{\Theta\left(q^{+}-k{}^{+}\right)\Theta\left(k^{+}\right)}{4k^{+}\left(q^{+}-k^{+}\right)}\,\frac{i}{q^{-}-\frac{\boldsymbol{k}{}^{2}_{\perp}+m^{2}}{2k^{+}}-\frac{\left(\boldsymbol{q}_{\perp}-\boldsymbol{k}_{\perp}\right)^{2}+m^{2}}{2\left(q^{+}-k{}^{+}\right)}}.\nonumber 
\end{align}
The Heaviside step functions $\Theta(\dots)$ in Eq.~(\ref{eq:Cauchy_2})
restrict the range of the $k^{+}$ integration and arise from the
requirement that the integrand has poles on both sides of the real
$k^{-}$ axis to yield a non-zero result. Integrating over $\boldsymbol{k}_{\perp}$
subsequently yields the Macdonald functions $K_{0}$ and $K_{1}$,
so that the amplitude~(\ref{eq:M}) reduces to 
\begin{align}
\mathcal{A}^{a} & =\bar{u}_{h}(k_{1})\,\mathcal{M}^{c,c'}_{\text{frag}}S_{F}\left(K\right)\gamma^{+}u_{h'}(K)\times\label{eq:Mred}\\
 & \times\int d^{2}\mathbf{x}_{1}d^{2}\mathbf{x}_{2}\,e^{-i\boldsymbol{K}_{\perp}\cdot\mathbf{x}_{1}-i\mathbf{k}_{2\perp}\cdot\mathbf{x}_{2}}\,\Psi^{(\lambda,\,h',\bar{h})}_{g\to\bar{b}b}(\alpha,\mathbf{x}_{1}-\mathbf{x}_{2})\,\left[U(\mathbf{x}_{1})t^{a}U^{\dagger}(\mathbf{x}_{2})-\mathcal{U}_{ab}\left(\boldsymbol{b}_{12}\right)t^{b}\right]_{\bar{c},c'},\nonumber 
\end{align}
where $\lambda$ is the helicity of the incoming gluon, $\bar{c},\bar{h}$
are the color and helicity indices of the produced antiquark, and
$h',h$ are the helicities of the quark before and after the fragmentation.
The gluon splitting wave function $\Psi^{(\lambda,h,\bar{h})}_{g\to\bar{b}b}$
at leading order in $\alpha_{s}$ is given by~\citep{Bjorken:1970ah,Dosch:1996ss}:
\begin{align}
\Psi^{(\lambda,\,h,\bar{h})}_{g\to\bar{b}b}\left(\alpha,\,r\right) & =-\frac{i\sqrt{4\pi\alpha_{s}}}{\sqrt{2k^{+}\,2\left(q^{+}-k^{+}\right)}}\int\frac{d^{2}\boldsymbol{k}_{\perp}}{\left(2\pi\right)^{2}}e^{i\boldsymbol{k}_{\perp}\cdot\boldsymbol{r}}\left.\frac{\bar{u}_{h}\left(k^{+},\,\boldsymbol{k}_{\perp}\right)\slashed{\varepsilon}_{\lambda}(q)v_{\bar{h}}\left(q^{+}-k^{+},\,-\boldsymbol{k}_{\perp}\right)}{q^{-}-\frac{q^{+}\left(\boldsymbol{k}^{2}_{\perp}+m^{2}\right)}{2k^{+}\left(q^{+}-k^{+}\right)}}\right|_{q^{-}=0,\quad k^{+}=\alpha q^{+}}\\
 & =\frac{\sqrt{8\pi\alpha_{s}}\,}{2\pi}\left[-i\lambda e^{-i\lambda\phi_{r}}\left(\alpha\delta_{h,\lambda}\delta_{\bar{h},-\lambda}-(1-\alpha)\delta_{h,-\lambda}\delta_{\bar{h},\lambda}\right)mK_{1}\left(mr\right)+m\delta_{h,\lambda}\delta_{\bar{h},\lambda}K_{0}\left(mr\right)\right].\nonumber 
\end{align}
To obtain the complete scattering amplitude, we must add the contributions
where the shockwave interacts directly with the incoming gluon, as
well as the terms where the charmonia are produced from the antiquark.
This procedure yields the total amplitude: 
\begin{align}
\mathcal{A}^{a}_{{\rm tot}} & =\frac{1}{2\sqrt{\alpha\left(1-\alpha\right)}}\int d^{2}\mathbf{x}_{1}d^{2}\mathbf{x}_{2}\,\Psi^{(\lambda,\,h',\bar{h}')}_{g\to\bar{b}b}(\alpha,\mathbf{x}_{1}-\mathbf{x}_{2})\left[U(\mathbf{x}_{1})t^{a}U^{\dagger}(\mathbf{x}_{2})-\mathcal{U}_{ab}\left(\boldsymbol{b}_{12}\right)t^{b}\right]_{\bar{c}',c'}\times\label{eq:MAll}\\
 & \times\left[\delta_{\bar{h},\bar{h}'}\delta_{\bar{c},\bar{c}'}\bar{u}_{h'}\left(k_{1}\right)\,\mathcal{M}^{c',c}_{\text{frag}}\left(k_{1},p_{1},p_{2};K\right)S_{F}\left(K\right)\gamma^{+}u_{h}\left(K'\right)e^{-i\boldsymbol{K}_{\perp}\cdot\mathbf{x}_{1}-i\mathbf{k}_{2\perp}\cdot\mathbf{x}_{2}}+\right.\nonumber \\
 & \left.+\delta_{h,h'}\delta_{c,c'}\bar{v}_{\bar{h}}\left(K'\right)\gamma^{+}S_{F}\left(K\right)\,\mathcal{M}^{\bar{c},\bar{c}'}_{\text{frag}}\left(-k_{2},p_{1},p_{2};-K\right)v_{\bar{h}'}\left(k_{2}\right)e^{-i\boldsymbol{K}_{\perp}\cdot\mathbf{x}_{2}-i\mathbf{k}_{1\perp}\cdot\mathbf{x}_{1}}\right].\nonumber 
\end{align}
To obtain the cross section, we square the amplitude~(\ref{eq:MAll}),
sum over final-state color and spin indices, and average over the
initial-state spin and color configurations. In general, this procedure
leads to a convolution of the spin density matrices of the produced
$Q\bar{Q}$ system with the helicity-dependent matrix $\hat{\rho}_{\mathfrak{c},\mathfrak{c}}=\sum_{\mathfrak{j}}\mathcal{\hat{M}}^{\mathfrak{j}\mathfrak{c}}_{\text{frag}}\mathcal{\hat{M}}^{*\mathfrak{j}\mathfrak{c}'}_{\text{frag}}$,
where we introduced the joint index notation $\mathfrak{c}=(c,h)$,
$\mathfrak{c}'=(c',h')$ for color and helicity indices, and defined
a shorthand notation~\footnote{Due to $C$-parity of strong interactions, a similar relation holds
for the conjugate matrix projected onto spinors of antiquarks $v,\bar{v}$.}
\begin{equation}
\mathcal{\hat{M}}^{\mathfrak{c},\mathfrak{c}'}_{\text{frag}}\left(k_{1},\,p_{1},p_{2},K\right)=\bar{u}_{h}\left(k_{1}\right)\,\mathcal{M}^{c,c'}_{\text{frag}}\left(k_{1},p_{1},p_{2};K\right)S_{F}\left(K\right)\gamma^{+}u_{h'}\left(K'\right).
\end{equation}
In what follows, we focus on the dominant unpolarized, color-singlet
contributions, where the main contribution comes from $\mathfrak{i}=\mathfrak{i}'$.
We observe that the amplitude $\mathcal{M}^{ji}_{\text{frag}}$ relates
directly to the definition of the dihadron fragmentation function
in Eq.~(\ref{e:D1_quark_DiFF_def-2}) via 
\begin{equation}
\int\frac{d^{2}\boldsymbol{K}_{\perp}}{2\left(2\pi\right)^{3}}\frac{d^{2}\boldsymbol{R}_{\perp}}{2\left(2\pi\right)^{3}}\sum_{\mathfrak{i,j}}\left|\mathcal{\hat{M}}^{\mathfrak{j}\mathfrak{i}}_{\text{frag}}\left(k_{1},p_{1},p_{2};K\right)\right|^{2}=\frac{1-z}{z}D^{\mathcal{Q}_{1}\mathcal{Q}_{2}/q}_{1}\left(z,\zeta\right),\label{eq:MSq}
\end{equation}
where the integration variables $\boldsymbol{K}_{\perp}$ and $\boldsymbol{R}_{\perp}$
represent the transverse momentum of the leading quark and the relative
transverse momentum of the individual quarkonium states with respect
to the pair momentum $P=p_{1}+p_{2}$, respectively. The prefactor
on the right-hand side of Eq.~(\ref{eq:MSq}) arises from the different
phase-space factors for the production of a heavy quarkonium pair
versus a bare quark pair\footnote{Indeed, in heavy-quark pair production, the produced active quark
is on-shell and contributes a phase-space factor $dk'^{+}d^{2}\boldsymbol{k}'_{\perp}/[2k'^{+}(2\pi)^{3}]$,
whereas in the production of a heavy quark and a quarkonium pair,
the active quark contributes with $dk^{+}_{1}d^{2}\boldsymbol{k}_{1\perp}/[2k^{+}_{1}(2\pi)^{3}]$.
The transverse momenta $\boldsymbol{k}_{1\perp}$ and $\boldsymbol{k}_{\perp}'$
are related by Eq.~(\ref{eq:kRel}), so that $d^{2}\boldsymbol{k}_{1\perp}=d^{2}\boldsymbol{k}_{\perp}'$.
The longitudinal components are related by $k^{+}_{1}=(1-z)k'^{+}$,
and the factor of $z$ in the denominator comes from the definition~(\ref{e:D1_quark_DiFF_def-2}).}. The lower cutoff on the invariant mass $M_{12}$ discussed in Eq.~(\ref{eq:vrel})
translates into a restriction on the phase space: 
\begin{equation}
R^{2}_{\perp}\ge\frac{M^{2}_{{\rm min}}\left(1-\zeta^{2}\right)-2(1-\zeta)M^{2}_{1}-2(1+\zeta)M^{2}_{2}}{4}.
\end{equation}
The cross section for the dihadron fragmentation mechanism can then
be expressed as 
\begin{align}
\frac{d\sigma\left(g+p\to\mathcal{Q}_{1}+\mathcal{Q}_{2}+X\right)}{dy_{1}dy_{2}} & \approx\int^{1}_{z_{{\rm min}}}\frac{dz}{z^{2}}\left[D^{\mathcal{Q}_{1}\mathcal{Q}_{2}/q}_{1}\left(z,\zeta\right)+D^{\mathcal{Q}_{1}\mathcal{Q}_{2}/\bar{q}}_{1}\left(z,\zeta\right)\right]\frac{d\sigma\left(g+p\to Q+\bar{Q}+X\right)}{d\tilde{y}_{1}d\tilde{y}_{2}},\label{eq:sigma-1}
\end{align}
where the tilde in the right-hand side indicates that the rapidities
of the heavy quarks must be evaluated with respect to the parent quark,
being effectively shifted as $\tilde{y}_{i}=y_{i}+\ln(1/z)$, while
the parameter $\zeta$ is defined as $\zeta=\tanh(y_{1}-y_{2})$.
The lower bound $z_{\text{min}}$ in the integral of Eq.~(\ref{eq:sigma-1})
is determined by the requirement that the light-cone momentum of the
parent quark does not exceed the momentum of the incoming gluon. The
light-cone momentum fractions $x_{1}$ and $x_{2}$ carried by the
gluon from the projectile and exchanged with the target, respectively,
are given by 
\begin{align}
x^{({\rm DF})}_{1} & =\frac{q^{+}}{P^{+}_{1}}=\frac{M^{\perp}_{1}e^{y_{1}}}{\alpha\,z_{1}\sqrt{s}}=\frac{M^{\perp}_{2}e^{y_{2}}}{\alpha z_{2}\sqrt{s}},
\end{align}
\begin{equation}
x^{({\rm DF})}_{2}=\frac{\sum p^{-}_{{\rm out}}-\sum p^{-}_{{\rm in}}}{P^{-}_{2}}\approx\frac{\bar{z}M^{2}_{12}+zm^{2}(1-\alpha z)}{z\,(1-z)\,\alpha\,(1-\alpha)x_{1}s},
\end{equation}
where we neglected the transverse-momentum dependence in the dipole
amplitude due to its relatively mild effect on $x_{2}$ and mild dependence
of the dipole cross-section on $x_{2}$. In view of the $C$-parity
we may expect that the contributions of the quark and antiquark in~(\ref{eq:sigma-1})
will coincide. 

Similarly, we can calculate the cross section for quarkonium pair
production via the single-fragmentation mechanism. This process occurs
independently for each of the produced quarkonia and is described
by single-fragmentation functions $D^{\mathcal{Q}_{i}/q}_{1}(z_{i})$,
which represent the probability for a heavy quark with momentum fraction
$z_{i}$ to form a quarkonium state $\mathcal{Q}_{i}$. The corresponding
cross section is given by 
\begin{equation}
\frac{d\sigma\left(g+p\to\mathcal{Q}_{1}+\mathcal{Q}_{2}+X\right)}{dy_{1}dy_{2}}=\int^{1}_{\bar{z}_{{\rm min}}}\frac{dz_{1}}{z^{2}_{1}}\int^{1}_{\bar{z}_{{\rm min}}}\frac{dz_{2}}{z^{2}_{2}}D^{\mathcal{Q}_{1}/q}_{1}\left(z_{1}\right)D^{\mathcal{Q}_{2}/q}_{1}\left(z_{2}\right)\frac{d\sigma\left(g+p\to Q+\bar{Q}+X\right)}{d\bar{y}_{1}d\bar{y}_{2}},\label{eq:sigma}
\end{equation}
where $\bar{y}_{i}=y_{i}+\ln(1/z_{i})$ are the rapidities of the
heavy quarks prior to fragmentation. The lower bounds on the $z_{1}$
and $z_{2}$ integrals are determined by the condition that the sum
of the light-cone energies of the $Q$ and $\bar{Q}$ quarks does
not exceed the energy of the incoming gluon, 
\begin{equation}
q^{+}\ge\frac{1}{\sqrt{2}}\left(\frac{M^{\perp}_{1}e^{y_{1}}}{\,z_{1}}+\frac{M^{\perp}_{2}e^{y_{2}}}{z_{2}}\right).\label{eq:mom}
\end{equation}
The corresponding light-cone momentum fractions $x_{1}$ and $x_{2}$
are given by 
\begin{align}
x^{({\rm SF})}_{1} & =\frac{q^{+}}{P^{+}_{1}}=\frac{M^{\perp}_{1}e^{y_{1}}}{\alpha\,z_{1}\sqrt{s}}+\frac{M^{\perp}_{2}e^{y_{2}}}{\left(1-\alpha\right)z_{2}\sqrt{s}},
\end{align}
\begin{equation}
x^{({\rm SF})}_{2}=\frac{\sum p^{-}_{{\rm out}}-\sum p^{-}_{{\rm in}}}{P^{-}_{2}}=\frac{\Delta q^{-}}{P^{+}}\approx\frac{m^{2}}{x_{1}s\alpha\left(1-\alpha\right)}.
\end{equation}
Finally, we note that our analysis considers only the specific contributions
that can be clearly attributed to either single- or dihadron fragmentation.
The full cross section also includes various interference terms, such
as the interference between single- and double-fragmentation amplitudes,
or between dihadron fragmentation from different quarks (the second
and the third lines in~(\ref{eq:MAll})). While these interference
terms are suppressed when the quark and antiquark are well separated
in rapidity or transverse momentum, they can double the overall contribution
of the fragmentation mechanism in some regions. Since they do not
reduce to a standard fragmentation convolution, we will not discuss
them in what follows.

\section{Numerical estimates for the cross section}

\label{sec:Numer}

We focus on the production of $S$-wave charmonium pairs ($J/\psi$
and $\psi(2S)$ mesons), which have very clear experimental signatures
and whose color-singlet long-distance matrix elements (LDMEs) are
known with reasonable precision~\citep{Lansberg:2019adr,Baranov:2019lhm,Baranov:2016clx}.
As explained above, for the $p_{T}$-integrated cross sections, quadrupole
amplitudes do not contribute, and all nuclear interactions are described
by color-dipole scattering amplitudes. For definiteness, we employ
the bCGC parameterization of the color-dipole amplitude with fit parameters
from Ref.~\citep{RESH}, corresponding to a charm quark mass of $m_{c}\approx1.4$~GeV.
For the forward collinear PDFs, we use the ``HERAPDF20\_LO\_EIG''
dataset evaluated at the scale $\mu=2M_{J/\psi}$ (the same scale
is used for $\alpha_{s}$). Since our calculations are performed at
leading order, we do not consider the explicit DGLAP evolution of
the fragmentation functions, assuming that the LO expressions derived
in Sec.~\ref{subsec:Fragmentation-function} are evaluated at the
same scale $\mu$. We estimate the theoretical uncertainty due to
choice of the scale varying it by a factor of two.

\begin{figure}
\includegraphics[width=9cm]{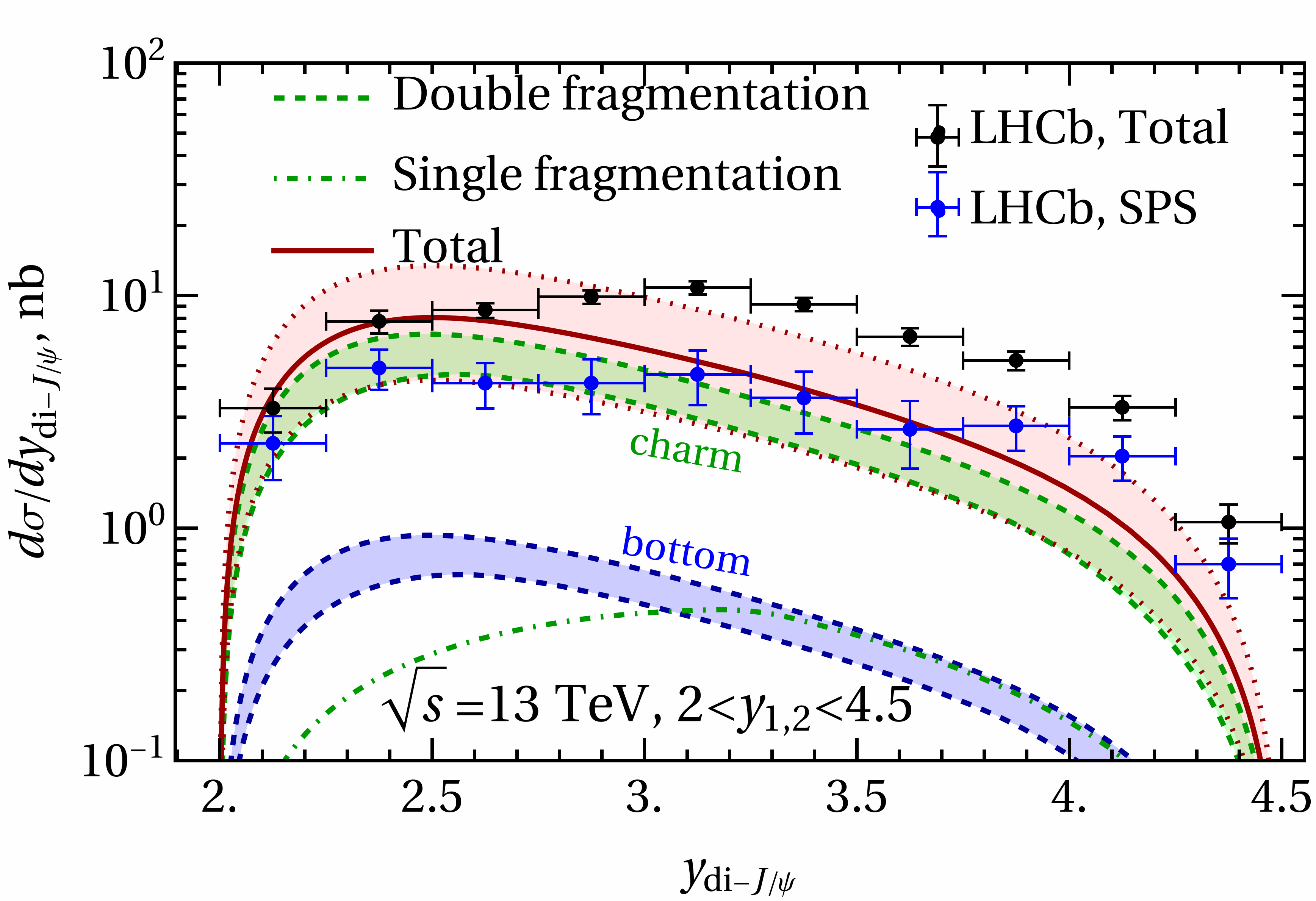}\includegraphics[width=9cm]{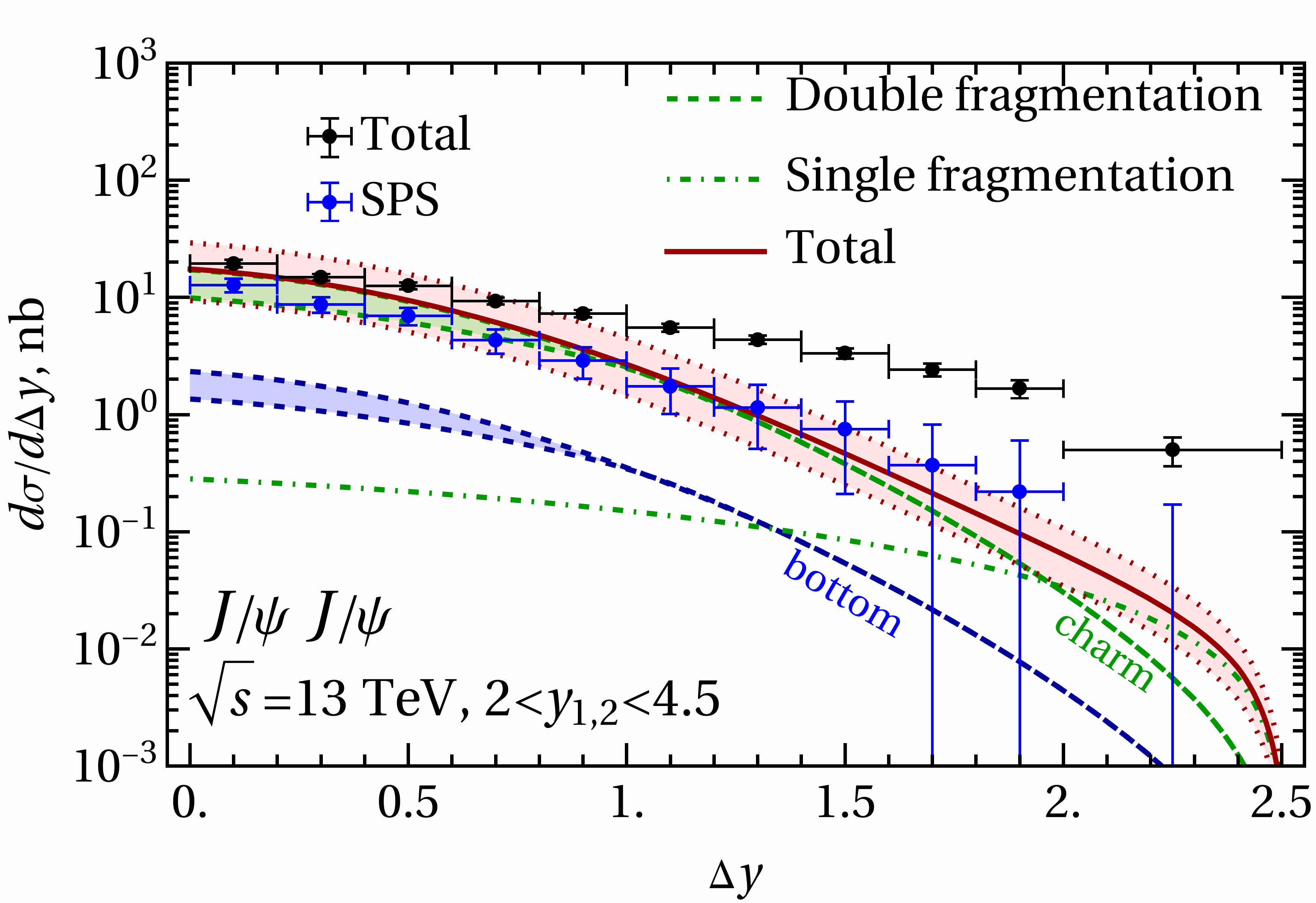}
\caption{ (Color online) Contributions of the single- and double-fragmentation
mechanisms to inclusive $J/\psi\,J/\psi$ production from heavy quarks.
For double fragmentation, the upper and lower boundaries of the bands
correspond to calculations using the full dihadron fragmentation function
and a version with an additional invariant mass cutoff $M^{\text{min}}_{12}=2M_{J/\psi}+\Lambda\approx7$~GeV,
respectively, so that the band width reflects the near-threshold contribution.
The green and blue bands represent the contributions of charm and
bottom DiFFs, respectively. The solid curve labeled \textquotedblleft Total\textquotedblright{}
includes both single- and double-fragmentation contributions from
charm and bottom quarks, with the colored band showing the uncertainty
from varying the scale $\mu$ by a factor of two and varying the collinear
gluon PDF parameterizations. Experimental data are from Ref.~\citep{LHCb:2023ybt}.}\label{fig:JJ}
\end{figure}

\begin{figure}
\includegraphics[width=9cm]{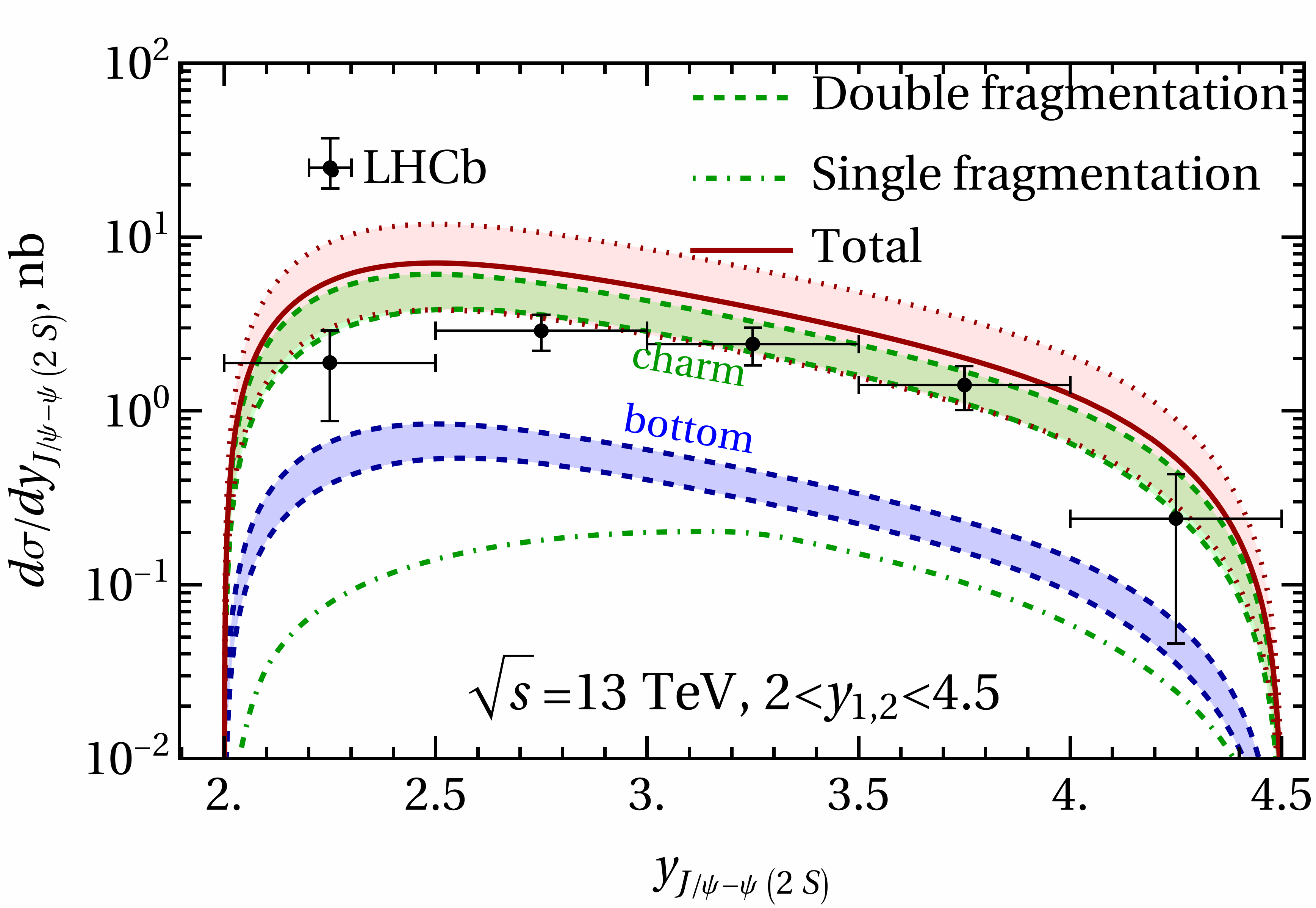}\includegraphics[width=9cm]{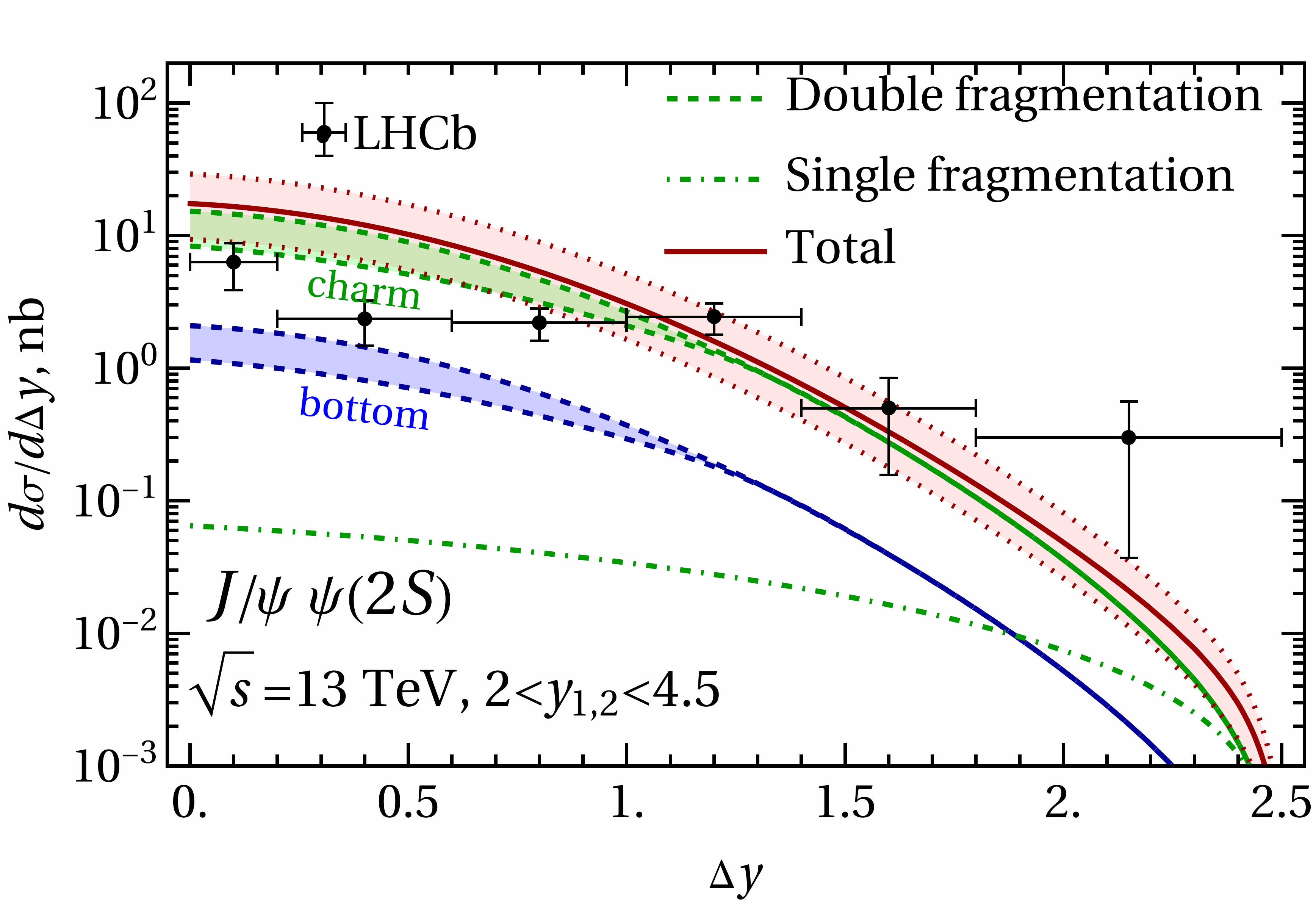}
\caption{ (Color online) Contributions of the single- and double-fragmentation
mechanisms to inclusive $J/\psi\,\psi(2S)$ production from heavy
$c$ and $b$ quarks. For double fragmentation, the upper and lower
boundaries of the bands correspond to calculations with the full dihadron
fragmentation function and with an additional invariant mass cutoff
$M^{\text{min}}_{12}=M_{J/\psi}+M_{\psi(2S)}+\Lambda\approx7.8$~GeV,
respectively. The green and blue bands correspond to the contributions
of charm and bottom DiFFs, respectively. The solid curve labeled \textquotedblleft Total\textquotedblright{}
shows the combined single- and double-fragmentation contributions
from charm and bottom quarks, with the band representing the uncertainty
from varying the scale $\mu$ by a factor of two and varying the collinear
gluon PDF parameterizations. Experimental data are from Ref.~\citep{LHCb:2023wsl}.}\label{fig:JP}
\end{figure}

In Figs.~\ref{fig:JJ}--\ref{fig:JP}, we compare the predicted
contributions of the fragmentation mechanisms with experimental data
from Refs.~\citep{LHCb:2023ybt,LHCb:2023wsl}. We verified that in
the relevant kinematics, the light-cone fractions satisfy $x_{1}\ge10^{-1}$
and $x_{2}\le10^{-2}$, justifying the use of the dilute-dense approximation.
In all plots the dot-dashed lines correspond to the independent (uncorrelated)
single fragmentation of each quark into a quarkonium state, which
remains negligible over the entire kinematic range. This suppression
occurs because single fragmentation either proceeds via small color-octet
LDMEs (for $b$ quarks) or requires the co-production of two additional
charmed quarks.

\begin{figure}
\includegraphics[width=9cm]{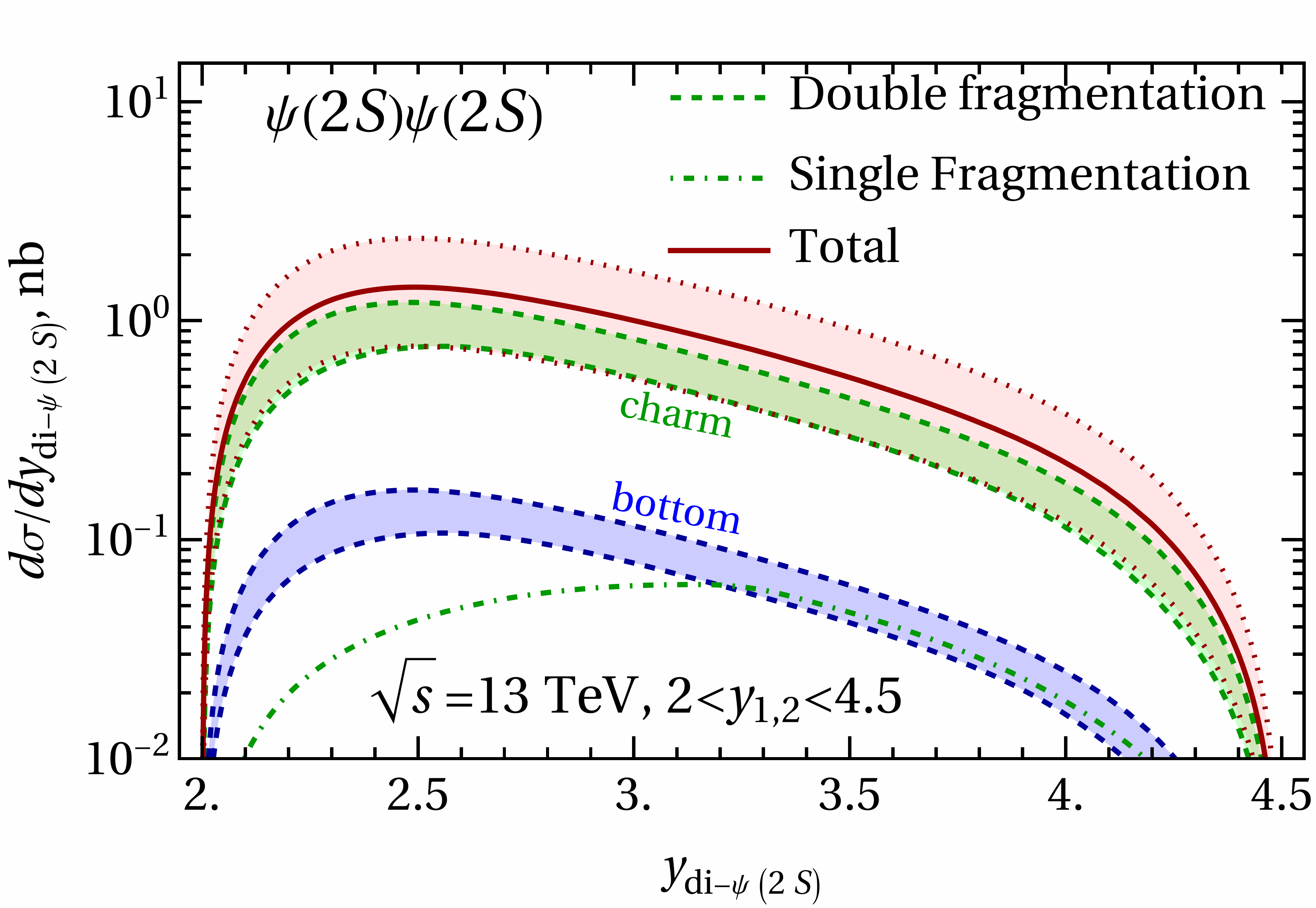}\includegraphics[width=9cm]{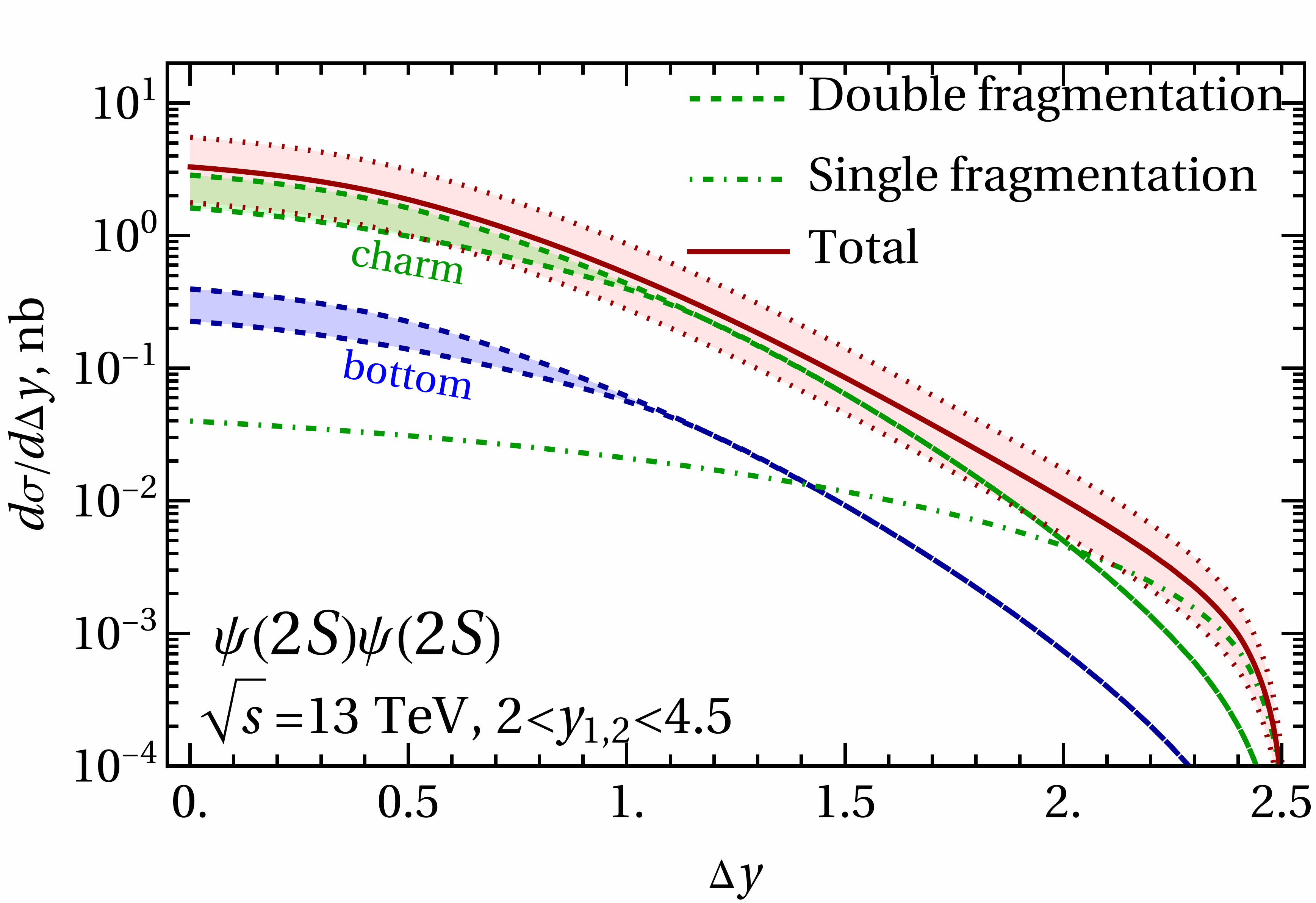}
\caption{ (Color online) Contributions of the single- and double-fragmentation
mechanisms to inclusive $\psi(2S)\,\psi(2S)$ production from heavy
$c$ and $b$ quarks. For double fragmentation, the upper and lower
boundaries of the bands represent the calculations with the full dihadron
fragmentation function and with an additional invariant mass cutoff
$M^{\text{min}}_{12}=2M_{\psi(2S)}+\Lambda\approx8.5$~GeV, respectively.
The green and blue bands represent the contributions of charm and
bottom DiFFs, respectively. The solid curve labeled \textquotedblleft Total\textquotedblright{}
represents the combined single- and double-fragmentation contributions
from charm and bottom quarks, with the band showing the uncertainty
from the scale $\mu$ variation and the collinear gluon PDF parameterizations.}\label{fig:PP}
\end{figure}

Since the perturbative treatment may lose reliability in the near-threshold
region, we provide predictions using both the full fragmentation function
and a version containing a cut on the invariant mass $M_{12}$, as
discussed below Eq.~(\ref{eq:M12}). The suppression of heavy-quark
contributions with increasing heavy-quark mass $m$ approximately
follows a power law $\sim1/m^{2\gamma}$, and is explained by the
suppression of the heavy-quark production cross section due to color
transparency (as we discussed earlier, the dihadron fragmentation
itself depends very weakly on the mass of the parent quark). The suppression
of the cross section at large rapidity separation $\Delta y$ arises
from the suppression of the dihadron fragmentation function at large
invariant masses $M_{12}$. The dependence on the rapidity $y$ is
determined by the interplay between the $x_{1}$ dependence of the
collinear PDF, the $x_{2}$ dependence of the dipole amplitude, and
the requirement that both charmonia fall within the acceptance of
the LHCb detector. Because the cross section decreases rapidly with
$p_{T}$, the fiducial cross section measured with an upper cutoff
$p_{T}<14$~GeV by the LHCb Collaboration practically coincides with
the $p_{T}$-integrated cross section studied in our framework. We
do not consider contributions from fragmentation of light quarks or
gluons, because there is no hard scale to justify a perturbative treatment
of the corresponding splitting subprocesses, such as $g\to q\bar{q}$
and $gg\to g$. Taken together, the charm and bottom contributions
are sizable and qualitatively reproduce the shape of the observed
rapidity dependence. In Fig.~\ref{fig:PP}, we also present predictions
for $\psi(2S)\,\psi(2S)$ production, which can be tested in future
experimental studies.

\section{Conclusions}

\label{sec:Conclusions}

In this manuscript, we evaluated the dihadron fragmentation function
and analyzed the contribution of fragmentation mechanisms to the inclusive
hadroproduction of heavy quarkonium pairs. We focused on the $p_{T}$-integrated
cross sections, which are controlled by forward dipole amplitudes.
Our results indicate that while single-quark fragmentation represents
only a small correction, dihadron fragmentation provides a significant
contribution that is on par with other production mechanisms. These
findings remain robust under variations of the renormalization scale,
parameterizations of the collinear gluon PDF, and the forward dipole
scattering amplitude, as well as under conservative lower cuts on
the invariant mass $M_{12}$ of the produced system.

Since at present our results indicate that the dihadron fragmentation
mechanism gives a significant contribution, the sum of this mechanism
together with direct SPS charmonia pair production (via $g+{\rm shockwave}\to\left(c\bar{c}\right)\left(c\bar{c}\right)\to\mathcal{Q}_{1}\mathcal{Q}_{2}$
subprocess), will exceed the experimental data, thus not leaving space
for the contribution of the Double Parton Scattering, and won't allow
extraction of the effective cross-section $\sigma_{{\rm eff}}$ that
encodes its magnitude. This tension underscores the importance of
next-to-leading-order (NLO) corrections. Indeed, it is well known
that NLO corrections can suppress the $Q\bar{Q}$ production cross
section by up to a factor of two~\citep{Caucal:2021ent,Caucal:2022ulg,Caucal:2023nci}.
A similar suppression in the dihadron fragmentation channel would
restore the space for DPS contributions, allowing for a meaningful
extraction of $\sigma_{\text{eff}}$. However, a systematic evaluation
of NLO corrections to the fragmentation mechanism requires computing
the corrections to both the single- and dihadron fragmentation functions,
as well as accounting for real emission, which is beyond the scope
of this work. Furthermore, such an analysis would require modeling
multipole (quadrupole and sextupole) operators, which are currently
poorly constrained phenomenologically.

Although we have focused on fragmentation in $pp$ collisions, extending
this framework to $pA$ collisions is straightforward. Since the fragmentation
time $t_{f}$ exceeds the interaction time $t_{h}$ by a factor of
$1/x_{2}\sim10^{2}$ in the considered kinematics, nuclear medium
effects should have only a minor influence on the fragmentation functions,
primarily affecting the forward dipole scattering amplitude~(\ref{eq:NS}).

\section*{Acknowledgments}

We thank our colleagues at UTFSM for encouraging discussions. This
research was partially supported by the project ``Proyecto ANID PIA/APOYO
AFB220004'' (Chile) and by ANID Fondecyt Regular Grants No.~1251322
and No.~1251975. This research was partially supported by the supercomputing
infrastructure of the NLHPC (ECM-02).

\appendix

\section{Evaluation of the diquarkonium fragmentation function}

\label{sec:cutoff-1}

In this appendix, we provide the technical details of the evaluation
of the dihadron fragmentation function. We perform the evaluations
in the heavy-quark mass limit and employ conventional NRQCD projectors.
We focus on $b$ quarks, for which the evaluation is the simplest.
Since the momenta of the quarkonia are shared equally between the
constituent charm quarks in the diagrams, both gluons carry the same
momentum $\ell=(P_{1}+P_{2})/2$ and are significantly off-shell,
namely $\ell^{2}=M^{2}_{12}/4$. The evaluation of the corresponding
diagram is straightforward and yields 
\begin{align}
D^{\mathcal{Q}_{1}\mathcal{Q}_{2}/q}_{1}(z,\,\zeta) & =C^{2}_{F}\int\frac{d^{2}\boldsymbol{K}_{\perp}}{2k'^{+}(2\pi)^{3}}\frac{d^{2}\boldsymbol{R}_{\perp}}{(2\pi)^{2}}\,V_{\mu\nu}V^{*}_{\alpha\beta}D_{\mu\mu'}(\ell)D_{\alpha\alpha'}(\ell)D_{\beta\beta'}(\ell)D_{\nu\nu'}(\ell)\times\\
 & \times\left.\text{Tr}\left(\gamma_{+}S(K)\gamma^{\mu'}S(K-\ell/2)i\gamma^{\nu'}\left(\slashed{K}-\slashed{P}_{1}-\slashed{P}_{2}+m\right)\gamma^{\alpha'}S(K-\ell/2)\gamma^{\beta'}S(K)\right)\right|_{K^{+}=P^{+}_{h}/z},\nonumber 
\end{align}
where the functions $V_{\mu\nu}$ correspond to the amplitude of the
$gg\to\mathcal{Q}_{1}\mathcal{Q}_{2}$ subprocess (the upper fermion
loop in the first diagram in Fig.~\ref{fig:diags_Fragmentation_Charm}),
$D_{\mu\mu'}$ are the gluon propagators, and $K^{\mu}$ is the momentum
of the parent quark, whose minus-component $K^{-}$ is fixed by the
requirement that the final-state quark is on-shell,
\begin{equation}
K^{-}=P^{-}_{1}+P^{-}_{2}+z\frac{\boldsymbol{k}^{2}_{1,\perp}+m^{2}}{2P^{+}_{h}\left(1-z\right)}.
\end{equation}
To evaluate the functions $V_{\mu\nu}$, we a need to project the
$Q\bar{Q}$ pairs onto states with definite color and spin. According
to potential models and NRQCD, the dominant Fock states of the $J/\psi$
and $\psi(2S)$ are the color-singlet $Q\bar{Q}$ pairs in the $^{3}S^{[1]}_{1}$
state. As discussed in Refs.~\citep{Cho:1995ce,Cho:1995vh,DVMPcc1},
the projectors onto color-singlet and color-octet states are given
respectively by 
\begin{align}
\left(\mathcal{P}^{[1]}\right)_{ij}=\frac{\delta_{ij}}{\sqrt{N_{c}}},\qquad\left(\mathcal{P}^{[8]}_{b}\right)_{ij} & =\sqrt{2}\,\left(t^{b}\right)_{ij},\quad b=1,\dots,8.
\end{align}
The projections onto a state with definite total spin $S$ and its
projection $S_{z}$ can be found using the appropriate Clebsch-Gordan
coefficients~\citep{Cho:1995ce,Cho:1995vh,DVMPcc1}: 
\begin{align}
\hat{\mathcal{P}}_{SS_{z}} & =\sum_{s_{1},\,s_{2}}\left\langle \frac{1}{2}s_{1}\frac{1}{2}s_{2}\bigg|1S_{z}\right\rangle v\left(\frac{P}{2}-\mathfrak{q},\,s_{2}\right)\bar{u}\left(\frac{P}{2}+\mathfrak{q},\,s_{1}\right)=\\
 & =\frac{-1}{2\sqrt{2}}\left(\frac{\slashed{P}}{2}-\slashed{\mathfrak{q}}-m_{c}\right)\slashed{\varepsilon}_{\mathcal{Q}}(P)\left(\frac{\slashed{P}}{2}+\slashed{\mathfrak{q}}+m_{c}\right),\nonumber 
\end{align}
where $P$ is the momentum of the produced quarkonium, $\mathfrak{q}\approx0$
is the momentum of relative motion of the quarks inside the quarkonium,
and $\varepsilon_{\mathcal{Q}}$ is the polarization vector of the
vector meson. Combining these projectors with the proper color-singlet
LDMEs and neglecting the relative momentum $\mathfrak{q}$, we obtain
the effective heavy-quark projectors onto the $J/\psi$ and $\psi(2S)$
states: {\small
\begin{align}
\left(\hat{\mathcal{P}}^{[1]}_{\mathcal{Q}}\right)_{ij} & \approx\sqrt{\frac{\left\langle \mathcal{O}^{[1]}_{\mathcal{Q}}\right\rangle }{m_{c}}}\frac{\delta_{ij}}{4N_{c}}\slashed{\varepsilon}_{\mathcal{Q}}(P)\left(\frac{\slashed{P}}{2}+m_{c}\right)\label{eq:PJPsi}
\end{align}
}where $\langle\mathcal{O}^{[1]}_{\mathcal{Q}}\rangle\equiv\langle\mathcal{O}^{[1]}_{\mathcal{Q}}(^{3}S^{[1]}_{1})\rangle$
are the corresponding color-singlet long-distance matrix elements
for $J/\psi$ and $\psi(2S)$ mesons. Phenomenological estimates,
for example, those based on the leptonic decay width of $J/\psi\to e^{+}e^{-}$,
suggest that $\langle\mathcal{O}^{[1]}_{J/\psi}(^{3}S^{[1]}_{1})\rangle\approx1.16\,\text{GeV}^{3}$
and $\langle\mathcal{O}^{[1]}_{\psi(2S)}(^{3}S^{[1]}_{1})\rangle\approx0.76\,\text{GeV}^{3}$~\citep{Braaten:2002fi,Lansberg:2019adr}.

Using these projectors and taking the traces over Dirac and color
indices, we obtain the functions $V_{\mu\nu}$: 
\begin{align}
V_{\mu\nu} & =\text{Tr}\left[\hat{\mathcal{P}}^{[1]}_{\mathcal{Q}_{1}}\gamma_{\mu}\hat{\mathcal{P}}^{[1]}_{\mathcal{Q}_{2}}\gamma_{\nu}\right]=\frac{\sqrt{\left\langle \mathcal{O}^{[1]}_{\mathcal{Q}_{1}}\right\rangle \left\langle \mathcal{O}^{[1]}_{\mathcal{Q}_{2}}\right\rangle }}{16\,m_{c}N_{c}}\left[g^{\mu\nu}\left(4\left({P_{1}}\cdot\varepsilon_{2}\right)\left({P_{2}}\cdot\varepsilon_{1}\right)+2\left(\left(M_{1}-M_{2}\right)^{2}-M^{2}_{12}\right)\left(\varepsilon_{1}\cdot\varepsilon_{2}\right)\right)\right.\\
 & -2\left(\left(M_{1}-M_{2}\right)^{2}-M^{2}_{12}\right)\left({\varepsilon}^{\mu}_{1}{\varepsilon}^{\nu}_{2}+{\varepsilon}^{\nu}_{1}{\varepsilon}^{\mu}_{2}\right)-4\left({\varepsilon}^{\mu}_{1}{P_{2}}^{\nu}\left({P_{1}}\cdot\varepsilon_{2}\right)+{\varepsilon}^{\mu}_{2}{P_{1}}^{\nu}\left({P_{2}}\cdot\varepsilon_{1}\right)\right)\nonumber \\
 & +\left.4\left(\varepsilon_{1}\cdot\varepsilon_{2}\right)\left({P_{2}}^{\mu}{P_{1}}^{\nu}+{P_{1}}^{\mu}{P_{2}}^{\nu}\right)-4\left({P_{2}}^{\mu}\left({\varepsilon}^{\nu}_{1}\left({P_{1}}\cdot\varepsilon_{2}\right)\right)+{P_{1}}^{\mu}\left({\varepsilon}^{\nu}_{2}\left({P_{2}}\cdot\varepsilon_{1}\right)\right)\right)\right]\nonumber 
\end{align}
The evaluation of the remaining traces is straightforward but tedious,
and was performed using FeynCalc version 10.2~\citep{FeynCalc1,FeynCalc2}.
The summation over the polarizations of the final-state quarkonia
was performed using the standard identity 
\begin{equation}
\sum_{a}\varepsilon^{(a)}_{\alpha}\varepsilon^{(a)*}_{\beta}=g_{\alpha\beta}-\frac{P_{\alpha}P_{\beta}}{M^{2}}\label{eq:sum}
\end{equation}

The final result can be represented in a compact form as 
\begin{align}
D^{\mathcal{Q}_{1}\mathcal{Q}_{2}/q}_{1}\left(z,\zeta\right)= & \Gpre\left[\hat{c}_{0}+\hat{c}_{A}\,\log\big(1+\varrho\big)+\hat{c}_{B}\,\log\big(1-\tfrac{z}{2}+\varrho\big)+\hat{c}_{U}\,\log\big(1-\tfrac{z}{2}\big)\right],\label{eq:master}
\end{align}
where we introduced the shorthand notations 
\begin{equation}
\varrho=\frac{m^{2}z^{2}}{\Mmin^{2}\,(1-z)},\qquad\Gpre=\frac{2\,C^{2}_{F}\alpha^{4}_{s}\left\langle \mathcal{O}^{[1]}_{\mathcal{Q}_{1}}\right\rangle \left\langle \mathcal{O}^{[1]}_{\mathcal{Q}_{2}}\right\rangle (1-z)\,(1-\zeta^{2})}{15\,N^{2}_{c}\,m^{8}m^{2}_{c}\,\Mmin^{12}\,z^{12}},\label{eq:prefactor}
\end{equation}
and $\Mmin^{2}=2\left((1-\zeta)M^{2}_{1}+(1+\zeta)M^{2}_{2}\right)/(1-\zeta^{2})$
is the minimal value of the invariant mass $M_{12}$ of the quarkonium
pair at a given $\zeta$. The coefficients $\hat{c}_{0},\hat{c}_{A},\hat{c}_{B},\hat{c}_{U}$
are polynomial functions of the variables $z$ and $\zeta$, whose
explicit forms are specified below. In the limit $z\to0$, the function
$D^{\mathcal{Q}_{1}\mathcal{Q}_{2}/q}_{1}(z,\zeta)$ increases only
as $\mathcal{O}(1/z)$ due to algebraic cancellations of the leading
poles. This behavior is explained by the increase of the phase space
volume with the energy of the parent quark $q^{+}=P^{+}_{h}/z$, suppressed
by an additional factor of $z$ in the definition~(\ref{e:D1_quark_DiFF_def-2}).

On general grounds, we expect the fragmentation function $D^{\mathcal{Q}_{1}\mathcal{Q}_{2}/q}_{1}(z,\zeta)$
should be invariant under the permutation of the two quarkonium states;
specifically, it should remain unchanged under the simultaneous exchange
of their masses $M_{1}\leftrightarrow M_{2}$ and the swap of the
kinematic variables $\zeta\to-\zeta$. To demonstrate that the resulting
expression possesses the desired symmetry and to present the explicit
expressions for the functions $\hat{c}_{0},\hat{c}_{A},\hat{c}_{B},\hat{c}_{U}$
in a compact form, we introduce auxiliary variables that respect this
symmetry. First, we define the constants 
\begin{equation}
S=M^{2}_{1}+M^{2}_{2},\qquad P=M_{1}M_{2},\qquad D=M^{2}_{1}-M^{2}_{2},\label{eq:SPD}
\end{equation}
that satisfy the identity $D^{2}=S^{2}-4P^{2}$. Since $D$ is odd
under the permutation of the two mesons, it appears in the final result
multiplied by odd powers of $\zeta$, whereas the symmetric combinations
$S$ and $P$ are multiplied by even powers of $\zeta$.

The explicit expressions for the coefficients $\hat{c}_{0},\hat{c}_{A},\hat{c}_{B},\hat{c}_{U}$
are given below: 
\begin{align}
\hat{c}_{0}={} & -384m^{2}\Mmin^{10}Q_{1}z^{3}+2304m^{2}\Mmin^{10}Q_{1}z^{4}\\
 & \quad+z^{5}\Big(-96m^{2}\Mmin^{10}\left(580D\zeta P+136D\zeta S+852\zeta^{2}P^{2}-580P^{2}-213\zeta^{2}S^{2}-68S^{2}\right)\nonumber \\
 & \qquad-384m^{4}\Mmin^{8}\Big(8P^{2}+S^{2}-8D\zeta P-2D\zeta S-12\zeta^{2}P^{2}+3\zeta^{2}S^{2}\nonumber \\
 & \qquad\qquad-2\Mmin^{2}\left(33D\zeta^{3}-9D\zeta-24\zeta^{2}P+16P-21\zeta^{2}S+S\right)\Big)\Big)\nonumber \\
 & \quad+z^{6}\Big(144m^{2}\Mmin^{10}\left(768D\zeta P+154D\zeta S+1076\zeta^{2}P^{2}-768P^{2}-269\zeta^{2}S^{2}-77S^{2}\right)\nonumber \\
 & \qquad+48m^{4}\Mmin^{8}\Big(304P^{2}+38S^{2}-304D\zeta P-76D\zeta S-456\zeta^{2}P^{2}+114\zeta^{2}S^{2}\nonumber \\
 & \qquad\qquad+\Mmin^{2}\left(-2388D\zeta^{3}+644D\zeta+1744\zeta^{2}P-1136P+1516\zeta^{2}S-76S\right)\Big)\Big)\nonumber \\
 & \quad+z^{7}\Big(-120m^{2}\Mmin^{10}\Big(-1300P^{2}-98S^{2}+1300D\zeta P+196D\zeta S+1692\zeta^{2}P^{2}-423\zeta^{2}S^{2}\Big)\nonumber \\
 & \qquad-16m^{4}\Mmin^{8}\Big(2064P^{2}+231S^{2}-2064D\zeta P-462D\zeta S-2988\zeta^{2}P^{2}+747\zeta^{2}S^{2}\nonumber \\
 & \qquad\qquad+\Mmin^{2}\left(-14052D\zeta^{3}+3836D\zeta+10876\zeta^{2}P-6724P+8614\zeta^{2}S-474S\right)\Big)\nonumber \\
 & \qquad+8m^{6}\Mmin^{6}\Big(384P^{2}+48S^{2}-384D\zeta P-96D\zeta S-576\zeta^{2}P^{2}+144\zeta^{2}S^{2}\nonumber \\
 & \qquad\qquad-96\Mmin^{2}\left(33D\zeta^{3}-9D\zeta-24\zeta^{2}P+16P-21\zeta^{2}S+S\right)\nonumber \\
 & \qquad\qquad+\left(5940\zeta^{4}-7560\zeta^{2}+2100\right)\Mmin^{4}\Big)\Big)\nonumber \\
 & +z^{8}\Big(12m^{2}\Mmin^{10}\Big(-13508P^{2}-640S^{2}+13508D\zeta P+1280D\zeta S+16068\zeta^{2}P^{2}-4017\zeta^{2}S^{2}\Big)\nonumber \\
 & \qquad+8m^{4}\Mmin^{8}\Big(5820P^{2}+498S^{2}-5820D\zeta P-996D\zeta S-7812\zeta^{2}P^{2}+1953\zeta^{2}S^{2}\nonumber \\
 & \qquad\qquad+\Mmin^{2}\Big(-15740P-1080S+8950D\zeta+27620\zeta^{2}P+17660\zeta^{2}S-31470D\zeta^{3}\Big)\Big)\nonumber \\
 & \qquad-4m^{6}\Mmin^{6}\Big(2688P^{2}+336S^{2}-2688D\zeta P-672D\zeta S-4032\zeta^{2}P^{2}+1008\zeta^{2}S^{2}\nonumber \\
 & \qquad\qquad+\Mmin^{2}\Big(-9792P-672S+5568D\zeta+15168\zeta^{2}P+13152\zeta^{2}S-20736D\zeta^{3}\Big)\nonumber \\
 & \qquad\qquad+\left(37260\zeta^{4}-46200\zeta^{2}+12300\right)\Mmin^{4}\Big)\Big)\nonumber \\
 & \quad+z^{9}\Big(-18m^{2}\Mmin^{10}\Big(-7004P^{2}-178S^{2}+7004D\zeta P+356D\zeta S+7716\zeta^{2}P^{2}-1929\zeta^{2}S^{2}\Big)\nonumber \\
 & \qquad-8m^{4}\Mmin^{8}\Big(5484P^{2}+285S^{2}-5484D\zeta P-570D\zeta S-6624\zeta^{2}P^{2}+1656\zeta^{2}S^{2}\nonumber \\
 & \qquad\qquad+\Mmin^{2}\Big(-12080P-675S+6715D\zeta+23600\zeta^{2}P+10445\zeta^{2}S-22245D\zeta^{3}\Big)\Big)\nonumber \\
 & \qquad+8m^{6}\Mmin^{6}\Big(2224P^{2}+224S^{2}-2224D\zeta P-448D\zeta S-3120\zeta^{2}P^{2}+780\zeta^{2}S^{2}\nonumber \\
 & \qquad\qquad+\Mmin^{2}\left(-13650D\zeta^{3}+3850D\zeta+11120\zeta^{2}P-6800P+8090\zeta^{2}S-450S\right)\nonumber \\
 & \qquad\qquad+\left(22245\zeta^{4}-25610\zeta^{2}+6005\right)\Mmin^{4}\Big)\nonumber \\
 & \qquad+2m^{8}\Mmin^{4}\Big(21504P^{2}+2688S^{2}-21504D\zeta P-5376D\zeta S-32256\zeta^{2}P^{2}+8064\zeta^{2}S^{2}\nonumber \\
 & \qquad\qquad+\Mmin^{2}\Big(-9216P-3456S+8064D\zeta+36864\zeta^{2}P+26496\zeta^{2}S-44928D\zeta^{3}\Big)\nonumber \\
 & \qquad\qquad+\left(26640\zeta^{4}-12960\zeta^{2}-4080\right)\Mmin^{4}\Big)\Big)\nonumber \\
 & \quad-2m^{2}\Mmin^{4}z^{10}\Big(-48\Mmin^{6}\left(740D\zeta P+26D\zeta S+792\zeta^{2}P^{2}-740P^{2}-198\zeta^{2}S^{2}-13S^{2}\right)\nonumber \\
 & \qquad+m^{2}\Mmin^{4}\Big(-14316P^{2}-318S^{2}+14316D\zeta P+636D\zeta S+15588\zeta^{2}P^{2}-3897\zeta^{2}S^{2}\nonumber \\
 & \qquad\qquad+\Mmin^{2}\Big(24244P+834S-12956D\zeta-55756\zeta^{2}P-14074\zeta^{2}S+41952D\zeta^{3}\Big)\Big)\nonumber \\
 & \qquad+4m^{4}\Mmin^{2}\Big(2200P^{2}+128S^{2}-2200D\zeta P-256D\zeta S-2712\zeta^{2}P^{2}+678\zeta^{2}S^{2}\nonumber \\
 & \qquad\qquad+\Mmin^{2}\left(-9237D\zeta^{3}+2921D\zeta+9406\zeta^{2}P-5314P+4534\zeta^{2}S-264S\right)\nonumber \\
 & \qquad\qquad+\left(12660\zeta^{4}-12070\zeta^{2}+1830\right)\Mmin^{4}\Big)\nonumber \\
 & \qquad+m^{6}\Big(48384P^{2}+6048S^{2}-48384D\zeta P-12096D\zeta S-72576\zeta^{2}P^{2}+18144\zeta^{2}S^{2}\nonumber \\
 & \qquad\qquad+\Mmin^{2}\Big(-14976P-7776S+15264D\zeta+77184\zeta^{2}P+53856\zeta^{2}S-92448D\zeta^{3}\Big)\nonumber \\
 & \qquad\qquad+\left(54180\zeta^{4}-25320\zeta^{2}-7260\right)\Mmin^{4}\Big)\Big)\nonumber \\
 & \quad+2m^{2}\Mmin^{2}z^{11}\Big(-54\Mmin^{8}\left(244D\zeta P+12D\zeta S+268\zeta^{2}P^{2}-244P^{2}-67\zeta^{2}S^{2}-6S^{2}\right)\nonumber \\
 & \qquad+m^{2}\Mmin^{6}\Big(-6660P^{2}-90S^{2}+6660D\zeta P+180D\zeta S+7020\zeta^{2}P^{2}-1755\zeta^{2}S^{2}\nonumber \\
 & \qquad\qquad+\Mmin^{2}\left(13992D\zeta^{3}-3676D\zeta-22436\zeta^{2}P+7004P-2774\zeta^{2}S+174S\right)\Big)\nonumber \\
 & \qquad+m^{8}\Big(15360P^{2}+1920S^{2}-15360D\zeta P-3840D\zeta S-23040\zeta^{2}P^{2}\nonumber \\
 & \qquad\qquad+5760\zeta^{2}S^{2}+2304\Mmin^{2}\left(-3D\zeta^{3}-D\zeta+4\zeta^{2}P+4P+\zeta^{2}S-S\right)\nonumber \\
 & \qquad\qquad+\left(2160\zeta^{4}+1440\zeta^{2}+2160\right)\Mmin^{4}\Big)\nonumber \\
 & \qquad+8m^{6}\Mmin^{2}\Big(4896P^{2}+504S^{2}-4896D\zeta P-1008D\zeta S-6912\zeta^{2}P^{2}\nonumber \\
 & \qquad\qquad+1728\zeta^{2}S^{2}+\Mmin^{2}\left(-7434D\zeta^{3}+762D\zeta+6492\zeta^{2}P-228P+4188\zeta^{2}S-648S\right)\nonumber \\
 & \qquad\qquad+\left(4245\zeta^{4}-1730\zeta^{2}-355\right)\Mmin^{4}\Big)\nonumber \\
 & \qquad+m^{4}\Mmin^{4}\Big(5072P^{2}+88S^{2}-5072D\zeta P-176D\zeta S-5424\zeta^{2}P^{2}+1356\zeta^{2}S^{2}\nonumber \\
 & \qquad\qquad+\Mmin^{2}\left(-13764D\zeta^{3}+4952D\zeta+18652\zeta^{2}P-9508P+4438\zeta^{2}S-198S\right)\nonumber \\
 & \qquad\qquad+\left(14445\zeta^{4}-7710\zeta^{2}-1395\right)\Mmin^{4}\Big)\Big)\nonumber \\
 & \quad-2m^{2}\Mmin^{2}z^{12}\Big(-30\Mmin^{8}\left(92D\zeta P+8D\zeta S+108\zeta^{2}P^{2}-92P^{2}-27\zeta^{2}S^{2}-4S^{2}\right)\nonumber \\
 & \qquad+m^{2}\Mmin^{6}\Big(-2004P^{2}-66S^{2}+2004D\zeta P+132D\zeta S+2268\zeta^{2}P^{2}-567\zeta^{2}S^{2}\nonumber \\
 & \qquad\qquad+\Mmin^{2}\left(3336D\zeta^{3}-508D\zeta-5588\zeta^{2}P+812P-542\zeta^{2}S+102S\right)\Big)\nonumber \\
 & \qquad+2m^{4}\Mmin^{4}\Big(832P^{2}+8S^{2}-832D\zeta P-16D\zeta S-864\zeta^{2}P^{2}+216\zeta^{2}S^{2}\nonumber \\
 & \qquad\qquad+\Mmin^{2}\left(-1518D\zeta^{3}+514D\zeta+2504\zeta^{2}P-1016P+266\zeta^{2}S-6S\right)\nonumber \\
 & \qquad\qquad+\left(1155\zeta^{4}+200\zeta^{2}-455\right)\Mmin^{4}\Big)\nonumber \\
 & \qquad+4m^{6}\Mmin^{2}\Big(3456P^{2}+144S^{2}-3456D\zeta P-288D\zeta S-4032\zeta^{2}P^{2}+1008\zeta^{2}S^{2}\nonumber \\
 & \qquad\qquad+\Mmin^{2}\left(-3078D\zeta^{3}-426D\zeta+3144\zeta^{2}P+1224P+1506\zeta^{2}S-186S\right)\nonumber \\
 & \qquad\qquad+\left(1635\zeta^{4}-270\zeta^{2}+115\right)\Mmin^{4}\Big)\nonumber \\
 & \qquad+m^{8}\Big(15360P^{2}+1920S^{2}-15360D\zeta P-3840D\zeta S-23040\zeta^{2}P^{2}\nonumber \\
 & \qquad\qquad+5760\zeta^{2}S^{2}+2304\Mmin^{2}\left(-3D\zeta^{3}-D\zeta+4\zeta^{2}P+4P+\zeta^{2}S-S\right)\nonumber \\
 & \qquad\qquad+\left(2160\zeta^{4}+1440\zeta^{2}+2160\right)\Mmin^{4}\Big)\Big)\nonumber \\
 & \quad+2m^{2}\Mmin^{4}z^{13}\Big(-9\Mmin^{6}\left(28D\zeta P+4D\zeta S+36\zeta^{2}P^{2}-28P^{2}-9\zeta^{2}S^{2}-2S^{2}\right)\nonumber \\
 & \qquad+m^{2}\Mmin^{4}\Big(-252P^{2}-18S^{2}+252D\zeta P+36D\zeta S+324\zeta^{2}P^{2}-81\zeta^{2}S^{2}\nonumber \\
 & \qquad\qquad+\Mmin^{2}\left(420D\zeta^{3}-40D\zeta-620\zeta^{2}P+20P-110\zeta^{2}S+30S\right)\Big)\nonumber \\
 & \qquad+m^{6}\Big(-96\left(16D\zeta P-2D\zeta S+12\zeta^{2}P^{2}-16P^{2}-3\zeta^{2}S^{2}+S^{2}\right)\nonumber \\
 & \qquad\qquad+120\Mmin^{2}\left(-3D\zeta^{3}-5D\zeta+8\zeta^{2}P+8P-\zeta^{2}S+S\right)+\left(120\zeta^{4}+400\zeta^{2}+120\right)\Mmin^{4}\Big)\nonumber \\
 & \qquad+m^{4}\Mmin^{2}\Big(336P^{2}+24S^{2}-336D\zeta P-48D\zeta S-432\zeta^{2}P^{2}+108\zeta^{2}S^{2}\nonumber \\
 & \qquad+\Mmin^{2}\left(-420D\zeta^{3}+40D\zeta+620\zeta^{2}P-20P+110\zeta^{2}S-30S\right)+\left(285\zeta^{4}+110\zeta^{2}-15\right)\Mmin^{4}\Big)\Big)\nonumber 
\end{align}

\begin{align}
\hat{c}_{U}={} & \Mmin^{12}\left[\left(5760z-768\right)Q_{1}-192Q_{2}z^{2}+96Q_{7}z^{3}+80Q_{3}z^{4}-8Q_{4}z^{5}+80Q_{5}z^{6}-4Q_{6}z^{7}\right]\\
 & \quad+3\Mmin^{12}z^{8}\Big(-98052P^{2}-2658S^{2}+98052D\zeta P+5316D\zeta S+108684\zeta^{2}P^{2}-27171\zeta^{2}S^{2}\nonumber \\
 & \qquad+m^{2}\left(8100D\zeta^{3}-12100D\zeta-10760\zeta^{2}P+26360P-2720\zeta^{2}S-1080S\right)\nonumber \\
 & \qquad+\left(161820\zeta^{4}-171000\zeta^{2}+31260\right)m^{4}\Big)\nonumber \\
 & \quad-15\Mmin^{12}z^{9}\Big(-12692P^{2}-262S^{2}+12692D\zeta P+524D\zeta S+13740\zeta^{2}P^{2}-3435\zeta^{2}S^{2}\nonumber \\
 & \qquad+m^{2}\left(-6396D\zeta^{3}+840D\zeta+8724\zeta^{2}P-1260P+2034\zeta^{2}S-210S\right)\nonumber \\
 & \qquad+\left(9654\zeta^{4}-9212\zeta^{2}+782\right)m^{4}\Big)\nonumber \\
 & \quad+\Mmin^{12}z^{10}\Big(-80568P^{2}-1836S^{2}+80568D\zeta P+3672D\zeta S\nonumber \\
 & \qquad+87912\zeta^{2}P^{2}-21978\zeta^{2}S^{2}\nonumber \\
 & \qquad+m^{2}\Big(-21920P-1320S+12280D\zeta+101600\zeta^{2}P+14600\zeta^{2}S-65400D\zeta^{3}\Big)\nonumber \\
 & \qquad+\left(13005\zeta^{4}-17910\zeta^{2}-2475\right)m^{4}\Big)\nonumber \\
 & \quad+5\Mmin^{12}z^{11}\Big(4392P^{2}+144S^{2}-4392D\zeta P-288D\zeta S-4968\zeta^{2}P^{2}\nonumber \\
 & \qquad+1242\zeta^{2}S^{2}+m^{2}\left(4584D\zeta^{3}-800D\zeta-7576\zeta^{2}P+1384P-796\zeta^{2}S+108S\right)\nonumber \\
 & \qquad+\left(1053\zeta^{4}+630\zeta^{2}+9\right)m^{4}\Big)\nonumber \\
 & \quad+\Mmin^{12}z^{12}\Big(-3516P^{2}-174S^{2}+3516D\zeta P+348D\zeta S+4212\zeta^{2}P^{2}\nonumber \\
 & \qquad-1053\zeta^{2}S^{2}+m^{2}\left(-4596D\zeta^{3}+628D\zeta+7448\zeta^{2}P-872P+872\zeta^{2}S-192S\right)\nonumber \\
 & \qquad+\left(-2085\zeta^{4}-1090\zeta^{2}+235\right)m^{4}\Big)\nonumber \\
 & \quad+\Mmin^{12}z^{13}\Big(252P^{2}+18S^{2}-252D\zeta P-36D\zeta S-324\zeta^{2}P^{2}+81\zeta^{2}S^{2}\nonumber \\
 & \qquad+m^{2}\left(420D\zeta^{3}-40D\zeta-620\zeta^{2}P+20P-110\zeta^{2}S+30S\right)+\left(285\zeta^{4}+110\zeta^{2}-15\right)m^{4}\Big)\nonumber 
\end{align}

\begin{align}
\hat{c}_{A}={} & \left(5760z-768\right)\Mmin^{12}Q_{1}-192\Mmin^{12}Q_{2}z^{2}\\
 & \quad+96\Mmin^{12}z^{3}\Big(-3196P^{2}-344S^{2}+3196D\zeta P+688D\zeta S+4572\zeta^{2}P^{2}\nonumber \\
 & \qquad-1143\zeta^{2}S^{2}+m^{2}\left(-3180D\zeta^{3}+860D\zeta+2320\zeta^{2}P-1520P+2020\zeta^{2}S-100S\right)\Big)\nonumber \\
 & \quad+160\Mmin^{12}z^{4}\Big(2508P^{2}+228S^{2}-2508D\zeta P-456D\zeta S-3420\zeta^{2}P^{2}\nonumber \\
 & \qquad+855\zeta^{2}S^{2}+m^{2}\left(4938D\zeta^{3}-1354D\zeta-3716\zeta^{2}P+2396P-3080\zeta^{2}S+156S\right)\nonumber \\
 & \qquad+\left(594\zeta^{4}-756\zeta^{2}+210\right)m^{4}\Big)\nonumber \\
 & \quad-32\Mmin^{12}z^{5}\Big(11112P^{2}+804S^{2}-11112D\zeta P-1608D\zeta S-14328\zeta^{2}P^{2}+3582\zeta^{2}S^{2}\nonumber \\
 & \qquad+m^{2}\Big(18280P+1095S-10235D\zeta-28840\zeta^{2}P-21475\zeta^{2}S+35895D\zeta^{3}\Big)\nonumber \\
 & \qquad+\left(13770\zeta^{4}-17220\zeta^{2}+4650\right)m^{4}\Big)\nonumber \\
 & \quad+160\Mmin^{12}z^{6}\Big(1296P^{2}+72S^{2}-1296D\zeta P-144D\zeta S-1584\zeta^{2}P^{2}\nonumber \\
 & \qquad+396\zeta^{2}S^{2}+m^{2}\left(6432D\zeta^{3}-1960D\zeta-5720\zeta^{2}P+3560P-3572\zeta^{2}S+180S\right)\nonumber \\
 & \qquad+\left(5271\zeta^{4}-6386\zeta^{2}+1631\right)m^{4}\Big)\nonumber \\
 & \quad-16\Mmin^{12}z^{7}\Big(4488P^{2}+192S^{2}-4488D\zeta P-384D\zeta S-5256\zeta^{2}P^{2}+1314\zeta^{2}S^{2}\nonumber \\
 & \qquad+m^{2}\Big(22816P+876S-12284D\zeta-37024\zeta^{2}P-18076\zeta^{2}S+36588D\zeta^{3}\Big)\nonumber \\
 & \qquad+\left(53655\zeta^{4}-61480\zeta^{2}+14085\right)m^{4}\Big)\nonumber \\
 & \quad+z^{8}\Big(-96\Mmin^{12}\left(116D\zeta P+8D\zeta S+132\zeta^{2}P^{2}-116P^{2}-33\zeta^{2}S^{2}-4S^{2}\right)\nonumber \\
 & \qquad-320m^{2}\Mmin^{12}\left(-651D\zeta^{3}+251D\zeta+754\zeta^{2}P-478P+274\zeta^{2}S-12S\right)\nonumber \\
 & \qquad+18480\left(27\zeta^{4}-28\zeta^{2}+5\right)m^{4}\Mmin^{12}\nonumber \\
 & \qquad+16m^{8}\Mmin^{4}\Big(-5760P^{2}-720S^{2}+5760D\zeta P+1440D\zeta S+8640\zeta^{2}P^{2}\nonumber \\
 & \qquad\qquad-2160\zeta^{2}S^{2}+960\Mmin^{2}\left(15D\zeta^{3}-3D\zeta-12\zeta^{2}P+4P-9\zeta^{2}S+S\right)\nonumber \\
 & \qquad\qquad+\left(-12600\zeta^{4}+10800\zeta^{2}-1080\right)\Mmin^{4}\Big)\Big)\nonumber \\
 & \quad+80m^{2}\Mmin^{4}z^{9}\Big(2\Mmin^{8}\left(-267D\zeta^{3}+127D\zeta+344\zeta^{2}P-248P+95\zeta^{2}S-3S\right)\nonumber \\
 & \qquad+\left(-2064\zeta^{4}+1777\zeta^{2}-109\right)m^{2}\Mmin^{8}\nonumber \\
 & \qquad+m^{6}\Big(2880P^{2}+360S^{2}-2880D\zeta P-720D\zeta S-4320\zeta^{2}P^{2}+1080\zeta^{2}S^{2}\nonumber \\
 & \qquad\qquad+\Mmin^{2}\left(-6624D\zeta^{3}+1248D\zeta+5376\zeta^{2}P-1536P+3936\zeta^{2}S-480S\right)\nonumber \\
 & \qquad\qquad+\left(5580\zeta^{4}-4536\zeta^{2}+396\right)\Mmin^{4}\Big)\Big)\nonumber \\
 & \quad+z^{10}\Big(-1280m^{2}\Mmin^{12}\left(-3D\zeta^{3}+2D\zeta+4\zeta^{2}P-4P+\zeta^{2}S\right)\nonumber \\
 & \qquad+80\left(363\zeta^{4}-218\zeta^{2}-69\right)m^{4}\Mmin^{12}\nonumber \\
 & \qquad-160m^{8}\Mmin^{4}\Big(1368P^{2}+144S^{2}-1368D\zeta P-288D\zeta S-1944\zeta^{2}P^{2}\nonumber \\
 & \qquad\qquad+486\zeta^{2}S^{2}+\Mmin^{2}\left(-2688D\zeta^{3}+448D\zeta+2336\zeta^{2}P-512P+1520\zeta^{2}S-192S\right)\nonumber \\
 & \qquad\qquad+\left(2133\zeta^{4}-1518\zeta^{2}+69\right)\Mmin^{4}\Big)\nonumber \\
 & \qquad-16m^{10}\Mmin^{2}\Big(9216P^{2}+1152S^{2}-9216D\zeta P-2304D\zeta S-13824\zeta^{2}P^{2}\nonumber \\
 & \qquad\qquad+3456\zeta^{2}S^{2}+1440\Mmin^{2}\left(-9D\zeta^{3}+D\zeta+8\zeta^{2}P+5\zeta^{2}S-S\right)\nonumber \\
 & \qquad\qquad+\left(7200\zeta^{4}-2880\zeta^{2}-480\right)\Mmin^{4}\Big)\Big)\nonumber \\
 & \quad+z^{11}\Big(720\left(-3\zeta^{4}+\zeta^{2}+2\right)m^{4}\Mmin^{12}\nonumber \\
 & \qquad-1920m^{10}\Mmin^{2}\Big(-96P^{2}-12S^{2}+96D\zeta P+24D\zeta S+144\zeta^{2}P^{2}-36\zeta^{2}S^{2}\nonumber \\
 & \qquad\qquad+\Mmin^{2}\left(117D\zeta^{3}-9D\zeta-108\zeta^{2}P-12P-63\zeta^{2}S+15S\right)+\left(-63\zeta^{4}+22\zeta^{2}+1\right)\Mmin^{4}\Big)\nonumber \\
 & \qquad+80m^{8}\Mmin^{4}\Big(1224P^{2}+72S^{2}-1224D\zeta P-144D\zeta S-1512\zeta^{2}P^{2}\nonumber \\
 & \qquad\qquad+378\zeta^{2}S^{2}+\Mmin^{2}\left(-1848D\zeta^{3}+224D\zeta+1888\zeta^{2}P-256P+904\zeta^{2}S-96S\right)\nonumber \\
 & \qquad\qquad+\left(1341\zeta^{4}-660\zeta^{2}-69\right)\Mmin^{4}\Big)\Big)\nonumber \\
 & \quad+z^{12}\Big(m^{12}\Big(-61440P^{2}-7680S^{2}+61440D\zeta P+15360D\zeta S+92160\zeta^{2}P^{2}\nonumber \\
 & \qquad\qquad-23040\zeta^{2}S^{2}-9216\Mmin^{2}\left(-3D\zeta^{3}-D\zeta+4\zeta^{2}P+4P+\zeta^{2}S-S\right)\nonumber \\
 & \qquad\qquad+\left(-8640\zeta^{4}-5760\zeta^{2}-8640\right)\Mmin^{4}\Big)\nonumber \\
 & \qquad-80m^{8}\Mmin^{4}\Big(216P^{2}-216D\zeta P-216\zeta^{2}P^{2}+54\zeta^{2}S^{2}\nonumber \\
 & \qquad\qquad+\Mmin^{2}\left(-264D\zeta^{3}+32D\zeta+352\zeta^{2}P-64P+88\zeta^{2}S\right)+\left(171\zeta^{4}-12\zeta^{2}-51\right)\Mmin^{4}\Big)\nonumber \\
 & \qquad-64m^{10}\Mmin^{2}\Big(1008P^{2}+72S^{2}-1008D\zeta P-144D\zeta S-1296\zeta^{2}P^{2}\nonumber \\
 & \qquad\qquad+324\zeta^{2}S^{2}+\Mmin^{2}\left(-855D\zeta^{3}-105D\zeta+870\zeta^{2}P+390P+420\zeta^{2}S-90S\right)\nonumber \\
 & \qquad\qquad+\left(435\zeta^{4}-70\zeta^{2}+55\right)\Mmin^{4}\Big)\Big)\nonumber \\
 & \quad+z^{13}\Big(80m^{8}\Mmin^{6}\left(-12D\zeta^{3}+8D\zeta+\left(9\zeta^{4}-3\zeta^{2}-6\right)\Mmin^{2}+16\zeta^{2}P-16P+4\zeta^{2}S\right)\nonumber \\
 & \qquad+m^{10}\Big(-384\Mmin^{2}\left(16D\zeta P-2D\zeta S+12\zeta^{2}P^{2}-16P^{2}-3\zeta^{2}S^{2}+S^{2}\right)\nonumber \\
 & \qquad\qquad+480\Mmin^{4}\left(-3D\zeta^{3}-5D\zeta+8\zeta^{2}P+8P-\zeta^{2}S+S\right)+\left(480\zeta^{4}+1600\zeta^{2}+480\right)\Mmin^{6}\Big)\Big)\nonumber 
\end{align}

\begin{align}
\hat{c}_{B}={} & \left(768-5760z\right)\Mmin^{12}Q_{1}+192\Mmin^{12}Q_{2}z^{2}-96\Mmin^{12}Q_{7}z^{3}-80\Mmin^{12}Q_{3}z^{4}\\
 & \quad+8\Mmin^{12}Q_{4}z^{5}-80\Mmin^{12}Q_{5}z^{6}+4\Mmin^{12}Q_{6}z^{7}\nonumber \\
 & \quad+z^{8}\Big(-9\Mmin^{12}\Big(-32684P^{2}-886S^{2}+32684D\zeta P+1772D\zeta S+36228\zeta^{2}P^{2}-9057\zeta^{2}S^{2}\Big)\nonumber \\
 & \qquad+60m^{2}\Mmin^{12}\left(-405D\zeta^{3}+605D\zeta+538\zeta^{2}P-1318P+136\zeta^{2}S+54S\right)\nonumber \\
 & \qquad-180\left(2697\zeta^{4}-2850\zeta^{2}+521\right)m^{4}\Mmin^{12}\nonumber \\
 & \qquad-3m^{8}\Mmin^{4}\Big(-30720P^{2}-3840S^{2}+30720D\zeta P+7680D\zeta S\nonumber \\
 & \qquad\qquad+46080\zeta^{2}P^{2}-11520\zeta^{2}S^{2}+5120\Mmin^{2}\left(15D\zeta^{3}-3D\zeta-12\zeta^{2}P+4P-9\zeta^{2}S+S\right)\nonumber \\
 & \qquad\qquad+\left(-67200\zeta^{4}+57600\zeta^{2}-5760\right)\Mmin^{4}\Big)\Big)\nonumber \\
 & \quad+15\Mmin^{4}z^{9}\Big(\Mmin^{8}\left(12692D\zeta P+524D\zeta S+13740\zeta^{2}P^{2}-12692P^{2}-3435\zeta^{2}S^{2}-262S^{2}\right)\nonumber \\
 & \qquad+6m^{2}\Mmin^{8}\left(-1066D\zeta^{3}+140D\zeta+1454\zeta^{2}P-210P+339\zeta^{2}S-35S\right)\nonumber \\
 & \qquad+2\left(4827\zeta^{4}-4606\zeta^{2}+391\right)m^{4}\Mmin^{8}\nonumber \\
 & \qquad+m^{8}\Big(-15360P^{2}-1920S^{2}+15360D\zeta P+3840D\zeta S+23040\zeta^{2}P^{2}-5760\zeta^{2}S^{2}\nonumber \\
 & \qquad\qquad+\Mmin^{2}\Big(8192P+2560S-6656D\zeta-28672\zeta^{2}P-20992\zeta^{2}S+35328D\zeta^{3}\Big)\nonumber \\
 & \qquad\qquad+\left(-29760\zeta^{4}+24192\zeta^{2}-2112\right)\Mmin^{4}\Big)\Big)\nonumber \\
 & \quad+z^{10}\Big(-54\Mmin^{12}\left(1492D\zeta P+68D\zeta S+1628\zeta^{2}P^{2}-1492P^{2}-407\zeta^{2}S^{2}-34S^{2}\right)\nonumber \\
 & \qquad-40m^{2}\Mmin^{12}\left(-1635D\zeta^{3}+307D\zeta+2540\zeta^{2}P-548P+365\zeta^{2}S-33S\right)\nonumber \\
 & \qquad-45\left(289\zeta^{4}-398\zeta^{2}-55\right)m^{4}\Mmin^{12}\nonumber \\
 & \qquad+160m^{8}\Mmin^{4}\Big(1368P^{2}+144S^{2}-1368D\zeta P-288D\zeta S-1944\zeta^{2}P^{2}\nonumber \\
 & \qquad\qquad+486\zeta^{2}S^{2}+\Mmin^{2}\left(-2688D\zeta^{3}+448D\zeta+2336\zeta^{2}P-512P+1520\zeta^{2}S-192S\right)\nonumber \\
 & \qquad\qquad+\left(2133\zeta^{4}-1518\zeta^{2}+69\right)\Mmin^{4}\Big)\nonumber \\
 & \qquad+m^{10}\Mmin^{2}\Big(147456P^{2}+18432S^{2}-147456D\zeta P-36864D\zeta S\nonumber \\
 & \qquad\qquad-221184\zeta^{2}P^{2}+55296\zeta^{2}S^{2}+23040\Mmin^{2}\left(-9D\zeta^{3}+D\zeta+8\zeta^{2}P+5\zeta^{2}S-S\right)\nonumber \\
 & \qquad\qquad+\left(115200\zeta^{4}-46080\zeta^{2}-7680\right)\Mmin^{4}\Big)\Big)\nonumber \\
 & \quad+z^{11}\Big(90\Mmin^{12}\left(244D\zeta P+16D\zeta S+276\zeta^{2}P^{2}-244P^{2}-69\zeta^{2}S^{2}-8S^{2}\right)\nonumber \\
 & \qquad+20m^{2}\Mmin^{12}\left(-1146D\zeta^{3}+200D\zeta+1894\zeta^{2}P-346P+199\zeta^{2}S-27S\right)\nonumber \\
 & \qquad-45\left(117\zeta^{4}+70\zeta^{2}+1\right)m^{4}\Mmin^{12}\nonumber \\
 & \qquad+1920m^{10}\Mmin^{2}\Big(-96P^{2}-12S^{2}+96D\zeta P+24D\zeta S+144\zeta^{2}P^{2}-36\zeta^{2}S^{2}\nonumber \\
 & \qquad\qquad+\Mmin^{2}\left(117D\zeta^{3}-9D\zeta-108\zeta^{2}P-12P-63\zeta^{2}S+15S\right)+\left(-63\zeta^{4}+22\zeta^{2}+1\right)\Mmin^{4}\Big)\nonumber \\
 & \qquad-80m^{8}\Mmin^{4}\Big(1224P^{2}+72S^{2}-1224D\zeta P-144D\zeta S-1512\zeta^{2}P^{2}\nonumber \\
 & \qquad\qquad+378\zeta^{2}S^{2}+\Mmin^{2}\left(-1848D\zeta^{3}+224D\zeta+1888\zeta^{2}P-256P+904\zeta^{2}S-96S\right)\nonumber \\
 & \qquad\qquad+\left(1341\zeta^{4}-660\zeta^{2}-69\right)\Mmin^{4}\Big)\Big)\nonumber \\
 & \quad+z^{12}\Big(-3\Mmin^{12}\left(1172D\zeta P+116D\zeta S+1404\zeta^{2}P^{2}-1172P^{2}-351\zeta^{2}S^{2}-58S^{2}\right)\nonumber \\
 & \qquad-4m^{2}\Mmin^{12}\left(-1149D\zeta^{3}+157D\zeta+1862\zeta^{2}P-218P+218\zeta^{2}S-48S\right)\nonumber \\
 & \qquad+5\left(417\zeta^{4}+218\zeta^{2}-47\right)m^{4}\Mmin^{12}\nonumber \\
 & \qquad+80m^{8}\Mmin^{4}\Big(216P^{2}-216D\zeta P-216\zeta^{2}P^{2}+54\zeta^{2}S^{2}\nonumber \\
 & \qquad\qquad+\Mmin^{2}\left(-264D\zeta^{3}+32D\zeta+352\zeta^{2}P-64P+88\zeta^{2}S\right)+\left(171\zeta^{4}-12\zeta^{2}-51\right)\Mmin^{4}\Big)\nonumber \\
 & \qquad+64m^{10}\Mmin^{2}\Big(1008P^{2}+72S^{2}-1008D\zeta P-144D\zeta S-1296\zeta^{2}P^{2}\nonumber \\
 & \qquad\qquad+324\zeta^{2}S^{2}+\Mmin^{2}\left(-855D\zeta^{3}-105D\zeta+870\zeta^{2}P+390P+420\zeta^{2}S-90S\right)\nonumber \\
 & \qquad\qquad+\left(435\zeta^{4}-70\zeta^{2}+55\right)\Mmin^{4}\Big)\nonumber \\
 & \qquad+m^{12}\Big(61440P^{2}+7680S^{2}-61440D\zeta P-15360D\zeta S-92160\zeta^{2}P^{2}\nonumber \\
 & \qquad\qquad+23040\zeta^{2}S^{2}+9216\Mmin^{2}\left(-3D\zeta^{3}-D\zeta+4\zeta^{2}P+4P+\zeta^{2}S-S\right)\nonumber \\
 & \qquad\qquad+\left(8640\zeta^{4}+5760\zeta^{2}+8640\right)\Mmin^{4}\Big)\Big)\nonumber \\
 & \quad+z^{13}\Big(9\Mmin^{12}\left(28D\zeta P+4D\zeta S+36\zeta^{2}P^{2}-28P^{2}-9\zeta^{2}S^{2}-2S^{2}\right)\nonumber \\
 & \qquad+10m^{2}\Mmin^{12}\left(-42D\zeta^{3}+4D\zeta+62\zeta^{2}P-2P+11\zeta^{2}S-3S\right)-5\left(57\zeta^{4}+22\zeta^{2}-3\right)m^{4}\Mmin^{12}\nonumber \\
 & \qquad+m^{10}\Mmin^{2}\Big(384\left(16D\zeta P-2D\zeta S+12\zeta^{2}P^{2}-16P^{2}-3\zeta^{2}S^{2}+S^{2}\right)\nonumber \\
 & \qquad\qquad-480\Mmin^{2}\left(-3D\zeta^{3}-5D\zeta+8\zeta^{2}P+8P-\zeta^{2}S+S\right)+\left(-480\zeta^{4}-1600\zeta^{2}-480\right)\Mmin^{4}\Big)\nonumber \\
 & \qquad-80m^{8}\Mmin^{6}\left(-12D\zeta^{3}+8D\zeta+\left(9\zeta^{4}-3\zeta^{2}-6\right)\Mmin^{2}+16\zeta^{2}P-16P+4\zeta^{2}S\right)\Big),\nonumber 
\end{align}

where we have introduced the shorthand notations $Q_{i}(\zeta)$ for
polynomials of the variable $\zeta$ that are invariant under the
permutation of the two quarkonia: 
\begin{align}
Q_{1}={} & \Big(-8P^{2}-S^{2}+8D\zeta P+2D\zeta S+12\zeta^{2}P^{2}-3\zeta^{2}S^{2}\Big),\label{eq:Q1}
\end{align}
\begin{align}
Q_{2}={} & \Big(-128m^{2}P-804P^{2}-8m^{2}S-96S^{2}+72D\zeta m^{2}+804D\zeta P+192D\zeta S\nonumber \\
 & \qquad+192\zeta^{2}m^{2}P+1188\zeta^{2}P^{2}+168\zeta^{2}m^{2}S-297\zeta^{2}S^{2}-264D\zeta^{3}m^{2}\Big),\label{eq:Q2}
\end{align}
\begin{align}
Q_{3}={} & \Big(420m^{4}+4792m^{2}P+4392P^{2}+312m^{2}S+387S^{2}-2708D\zeta m^{2}\nonumber \\
 & \qquad-4392D\zeta P-774D\zeta S-1512\zeta^{2}m^{4}-7432\zeta^{2}m^{2}P-5940\zeta^{2}P^{2}-6160\zeta^{2}m^{2}S\nonumber \\
 & \qquad+1485\zeta^{2}S^{2}+9876D\zeta^{3}m^{2}+1188\zeta^{4}m^{4}\Big),\label{eq:Q3}
\end{align}
\begin{align}
Q_{4}={} & \Big(18600m^{4}+72640m^{2}P+25812P^{2}+4260m^{2}S+1539S^{2}\nonumber \\
 & \qquad-40580D\zeta m^{2}-25812D\zeta P-3078D\zeta S-68880\zeta^{2}m^{4}-113920\zeta^{2}m^{2}P\nonumber \\
 & \qquad-31968\zeta^{2}P^{2}-84820\zeta^{2}m^{2}S+7992\zeta^{2}S^{2}+141780D\zeta^{3}m^{2}+55080\zeta^{4}m^{4}\Big),\label{eq:Q4}
\end{align}
\begin{align}
Q_{5}={} & \Big(3262m^{4}+6844m^{2}P-828P^{2}+306m^{2}S-81S^{2}-3728D\zeta m^{2}\nonumber \\
 & \qquad+828D\zeta P+162D\zeta S-12772\zeta^{2}m^{4}-10660\zeta^{2}m^{2}P+1152\zeta^{2}P^{2}-6634\zeta^{2}m^{2}S\nonumber \\
 & \qquad-288\zeta^{2}S^{2}+11964D\zeta^{3}m^{2}+10542\zeta^{4}m^{4}\Big),\label{eq:Q5}
\end{align}
\begin{align}
Q_{6}={} & \Big(56400m^{4}+77636m^{2}P-67932P^{2}+1566m^{2}S-2943S^{2}\nonumber \\
 & \qquad-40384D\zeta m^{2}+67932D\zeta P+5886D\zeta S-245560\zeta^{2}m^{4}-110684\zeta^{2}m^{2}P\nonumber \\
 & \qquad+79704\zeta^{2}P^{2}-52646\zeta^{2}m^{2}S-19926\zeta^{2}S^{2}+107988D\zeta^{3}m^{2}+213720\zeta^{4}m^{4}\Big),\label{eq:Q6}
\end{align}
\begin{align}
Q_{7}={} & \Big(-1520m^{2}P-3116P^{2}-100m^{2}S-334S^{2}+860D\zeta m^{2}+3116D\zeta P\nonumber \\
 & \qquad+668D\zeta S+2320\zeta^{2}m^{2}P+4452\zeta^{2}P^{2}+2020\zeta^{2}m^{2}S-1113\zeta^{2}S^{2}-3180D\zeta^{3}m^{2}\Big).\label{eq:Q7}
\end{align}

\section{Diquarkonium fragmentation function in the helicity basis}

\label{sec:HelicityBasis}

In this appendix, we provide the technical details of the evaluation
of the full set of Feynman diagrams shown in Fig.~\ref{fig:diags_Fragmentation_Charm},
which is needed to obtain the dihadron fragmentation function of the
charm quark. This evaluation is highly nontrivial because of the large
number of combinatorial pairwise combinations arising from the interference
between different diagrams in the amplitude and its conjugate, and
the necessity of evaluating traces containing up to twenty Dirac matrices,
which typically results in extremely lengthy analytical expressions.
Fortunately, this evaluation simplifies significantly in the helicity
basis. Since the initiating (parent) and final-state quarks can only
have two helicity states, we only need to determine the amplitudes
with and without a helicity flip of the active quark. Precisely, the
squared amplitude appearing in the integrand of Eq.~(\ref{eq:MSq})
can be represented as 
\begin{equation}
\sum_{\mathfrak{i,j}}\left|\mathcal{M}^{\mathfrak{j}\mathfrak{i}}_{\text{frag}}\left(k_{1},P_{1},P_{2};K\right)\right|^{2}=\frac{\text{Tr}\left[\gamma_{+}\left(\slashed{K}+m\right)\hat{d}^{\dagger}\left(\slashed{K}-\slashed{P}_{1}-\slashed{P}_{2}+m\right)\hat{d}\left(\slashed{K}+m\right)\right]}{\left(K^{2}-m^{2}+i0\right)^{2}}\label{eq:MSq2}
\end{equation}
where $\hat{d}$ denotes the sum of the contributions of the cut diagrams
shown in Fig.~\ref{fig:CGCBasic-1}. The expressions for $\hat{d}$
and $\hat{d}^{\dagger}$ also include the polarization vectors of
the produced quarkonia; for the unpolarized case, we sum over these
polarizations using the identity~(\ref{eq:sum}). Since the direct
evaluation of the trace on the right-hand side of Eq.~(\ref{eq:MSq2})
is mathematically challenging, we switch to the helicity representation.
Using the fact that the final-state quark momentum $K-P_{1}-P_{2}$
is on-shell, we may write 
\begin{equation}
\left(\slashed{K}-\slashed{P}_{1}-\slashed{P}_{2}+m\right)=\left\{ \begin{array}{c}
\sum_{h_{2}}u_{h_{2}}\left(K-P_{1}-P_{2}\right)\bar{u}_{h_{2}}\left(K-P_{1}-P_{2}\right),\qquad K^{+}>P^{+}_{1}+P^{+}_{2}\\
-\sum_{h_{2}}v_{h_{2}}\left(K-P_{1}-P_{2}\right)\bar{v}_{h_{2}}\left(K-P_{1}-P_{2}\right),\qquad K^{+}<P^{+}_{1}+P^{+}_{2}
\end{array}\right..
\end{equation}
Although the momentum $K$ is off-shell ($K^{2}\neq m^{2}$), we can
decompose the numerator $(\slashed{K}+m)$ into a sum over on-shell
spinors and an additional term that accounts for the off-shellness
of the quark: 
\begin{align}
\left(\slashed{K}+m\right) & =\left(K^{+}\gamma^{-}+\frac{\boldsymbol{K}^{2}_{\perp}+m^{2}}{2K^{+}}\gamma^{+}-\boldsymbol{K}_{\perp}\cdot\boldsymbol{\gamma}_{\perp}+m\right)+\gamma^{+}\left(K^{-}-\frac{\boldsymbol{K}^{2}_{\perp}+m^{2}}{2K^{+}}\right)=\\
 & =\theta\left(k'^{+}\right)\sum_{h_{1}}u_{h_{1}}\left(K'\right)\bar{u}_{h_{1}}\left(K'\right)-\theta\left(-K^{+}\right)\sum_{h_{1}}v_{h_{1}}\left(K'\right)\bar{v}_{h_{1}}\left(K'\right)+\gamma^{+}\left(K^{-}-\frac{\boldsymbol{K}^{2}_{\perp}+m^{2}}{2K^{+}}\right)\nonumber \\
 & K'=\left(K^{+},\,\frac{K^{2}_{\perp}+m^{2}}{2K^{+}},\,\boldsymbol{K}_{\perp}\right).
\end{align}
In the framework of light-cone perturbation theory~\citep{Lepage:1980fj,Brodsky:1997de},
the term proportional to $\gamma^{+}$ corresponds to the instantaneous
fermion propagator. Since $(\slashed{K}+m)$ in Eq.~(\ref{eq:MSq2})
is contracted with $\gamma^{+}$, the identity $\gamma^{+}\gamma^{+}=0$
ensures that this instantaneous contribution vanishes. Furthermore,
a cyclic permutation of the matrices within the trace, combined with
the identities~(\ref{eq:Id}) and (\ref{eq:S}), allows us to express
the squared amplitude of the $q\to q\mathcal{Q}_{1}\mathcal{Q}_{2}$
subprocess as a sum of helicity-flip and helicity-conserving contributions,
\begin{equation}
\sum_{\mathfrak{i,j}}\left|\mathcal{M}^{\mathfrak{j}\mathfrak{i}}_{\text{frag}}\left(k_{1},P_{1},P_{2};K\right)\right|^{2}=2K^{+}\sum_{h_{1}}\mathcal{M}^{*}_{h_{1},h_{1}}\,\mathcal{M}_{h_{1},h_{1}}+2K^{+}\sum_{h_{1}}\mathcal{M}^{*}_{h_{1},-h_{1}}\,\mathcal{M}_{-h_{1},h_{1}}.
\end{equation}
Using the algebraic identity 
\begin{align}
 & \left|\mathcal{M}_{++}\right|^{2}+\left|\mathcal{M}_{--}\right|^{2}+\left|\mathcal{M}_{+-}\right|^{2}+\left|\mathcal{M}_{-+}\right|^{2}=\\
 & =2\left[\left(\frac{\mathcal{M}_{++}+\mathcal{M}_{--}}{2}\right)^{2}+\left(\frac{\mathcal{M}_{++}-\mathcal{M}_{--}}{2}\right)^{2}+\left(\frac{\mathcal{M}_{+-}+\mathcal{M}_{-+}}{2}\right)^{2}+\left(\frac{\mathcal{M}_{+-}-\mathcal{M}_{-+}}{2}\right)^{2}\right],\nonumber 
\end{align}
and the properties of the light-cone Dirac spinors~\citep{Lepage:1980fj,Brodsky:1997de}:
\begin{align*}
\bar{u}_{h}\left(k_{1}\right)\gamma^{+}u_{h'}\left(k_{2}\right) & =2\sqrt{k^{+}_{1}k^{+}_{2}}\delta_{h,h'},\qquad\bar{u}_{h}\left(k_{1}\right)\gamma_{5}\gamma^{+}u_{h'}\left(k_{2}\right)=2\sqrt{k^{+}_{1}k^{+}_{2}}\delta_{h,h'}\left(-1\right)^{h+1/2},\\
\bar{u}_{h}\left(k_{1}\right)i\gamma_{2}\gamma^{+}u_{h'}\left(k_{2}\right) & =2\sqrt{k^{+}_{1}k^{+}_{2}}\delta_{h,-h'}\qquad\bar{u}_{h}\left(k_{1}\right)i\gamma_{5}\gamma_{2}\gamma^{+}u_{h'}\left(k_{2}\right)=2\sqrt{k^{+}_{1}k^{+}_{2}}\delta_{h,-h'}\left(-1\right)^{h+1/2},
\end{align*}
we obtain 
\begin{align}
\mathcal{M}_{++}+\mathcal{M}_{--} & =\frac{\text{Tr}\left[\,\hat{d}\left(\slashed{K}+m\right)\gamma^{+}\left(\slashed{K}-\slashed{P}_{1}-\slashed{P}_{2}+m\right)\right]}{2\sqrt{K^{+}\left(K^{+}-P^{+}_{1}-P^{+}_{2}\right)}},\label{eq:A}\\
\mathcal{M}_{++}-\mathcal{M}_{--} & =\frac{\text{Tr}\left[\,\hat{d}\left(\slashed{K}+m\right)\gamma_{5}\gamma^{+}\left(\slashed{K}-\slashed{P}_{1}-\slashed{P}_{2}+m\right)\right]}{2\sqrt{K^{+}\left(K^{+}-P^{+}_{1}-P^{+}_{2}\right)}},\label{eq:A1}
\end{align}
\begin{align}
\mathcal{M}_{+-}+\mathcal{M}_{-+} & =\frac{\text{Tr}\left[\,\hat{d}\left(\slashed{K}+m\right)i\gamma_{2}\gamma^{+}\left(\slashed{K}-\slashed{P}_{1}-\slashed{P}_{2}+m\right)\right]}{2\sqrt{K^{+}\left(K^{+}-P^{+}_{1}-P^{+}_{2}\right)}},\label{eq:A2}
\end{align}
\begin{align}
\mathcal{M}_{+-}-\mathcal{M}_{-+} & =\frac{\text{Tr}\left[\,\hat{d}\left(\slashed{K}+m\right)i\gamma_{5}\gamma_{2}\gamma^{+}\left(\slashed{K}-\slashed{P}_{1}-\slashed{P}_{2}+m\right)\right]}{2\sqrt{K^{+}\left(K^{+}-P^{+}_{1}-P^{+}_{2}\right)}}.\label{eq:A3}
\end{align}
The contributions~(\ref{eq:A1})--(\ref{eq:A3}) involve combinations
that are antisymmetric with respect to the permutation of the two
quarkonia, and for this reason are numerically suppressed in the $\boldsymbol{R}_{\perp}$-integrated
collinear fragmentation functions. While the following evaluation is straightforward, the explicit \textit{FeynCalc}-generated
expressions for the contribution of charm to the dihadron fragmentation function are very lengthy and can't be presented here due to space limitations. The full expressions can be provided on demand as a C/C++ code.

\end{document}